\documentclass[twocolumn,superscriptaddress,amsmath,amssymb,showpacs,floatfix,preprintnumbers]{revtex4-1}
\usepackage{amsmath}
\usepackage{amssymb}
\usepackage{graphicx}
\usepackage{color}
\usepackage{epsfig}
\usepackage{amsmath}
\usepackage{amssymb}
\usepackage{mathrsfs}
\graphicspath{ {images/} }
\usepackage{caption}
\usepackage{subcaption}
\usepackage{hyperref}

\DeclareMathAlphabet{\bi}{OML}{cmm}{b}{it}

\begin{document}
\title{Linear response function in the presence of elastic scattering: \\  plasmons in graphene and  the two-dimensional electron gas}
\author{M. Bahrami} 
\email{m$_$bahra@live.concordia.ca}
\author{P. Vasilopoulos}
\email{p.vasilopoulos@concordia.ca}
\ \\ 
\affiliation{Department of Physics, Concordia University, 7141 Sherbrooke Ouest, Montreal, Quebec, H4B 1R6, Canada}
\begin{abstract}
Within linear-response theory we derive a response function that thoroughly takes into account the influence of elastic scattering and is valid beyond the long-wavelength limit.  We apply the theo-\\ry to plasmons in graphene and the two-dimensional electron gas (2DEG), in the  random-phase ap-\\proximation, and for the former take into account  intraband and interband excitations. The scatte-\\ring-modified dispersion relation shows that below a critical scattering strength $\gamma_c$, simply related to  the plasmon frequency $\omega$, no plasmons are allowed.  The critical  strengths  $\gamma_c$ and the allowed $(\omega, q)$ plasmon spectra  for intraband and interband transitions in  graphene  are  different.   In both  graphene  and the 2DEG the strength $\gamma_c$ falls rapidly for small  $\omega$  but much   more slowly for large $\omega$. 

\end{abstract}

\maketitle
\section{Introduction}
Physical properties of a system such as conductivity,  susceptibility, etc., can be determined by studying  its  response to external probes, such as  electric or magnetic fields.  Many  properties in the classical regime could be explained by classical equations, such as the Boltzmann equation \cite{R-1}.  One  powerful tool to deal with many-body systems, which provides us with a wealth of information,  is linear response theory (LRT).  A classical LRT example is the Brownian motion \cite{R-2}, 
\cite{R-3}. To describe such properties of a system  quantum mechanically one can use the quantum master equation or the quantum linear response theory (QLRT)  introduced by  Kubo in the middle of last century \cite{R-4}. QLRT could provide  us with vital information about system properties at zero and finite temperature \cite{R-5}. For instance, static screening, effective interactions, collective modes, electron energy-loss spectra and Raman spectra  of a system can be obtained from its density-density response function (DDRF) \cite{R-6}.  One  major motivation in developing LRT was   finding the connection between fluctuations of a quantity about equilibrium and dissipation, which is the content of the fluctuation-dissipation theorem. Obviously it was of interest to have a quantum mechanical version of this theorem. 

Kubo's QLRT has been often criticized, see, e.g., Ref. \cite{R-1} and references cited therein. In particular, in developing QLRT Kubo 
assumed an adiabatic switching-on of the perturbation to avoid a divergence in an integration over time and satisfy causality. The relevant parameter though is regarded in the literature as a dissipative factor, see Eq. (3.42) in Ref.  \cite{R-6}. Moreover, to retrieve Drude's result  for the conductivity in the long-wavelength limit  usually one enters  this parameter in the final result phenomenologically, see, e.g., Refs. \cite{EX-1,EX-2,EX-3,EX-4,EX-5,EX-6,EX-7,EX-8,EX-9}. As  mentioned in Ref.  \cite{R-1}, the dissipation should  originate from some randomness in the system  and the latter should appear in the Hamiltonian. This randomness term, for example, could represent one-body or two-body  randomizing collisions \cite{R-7,R-1}. In Ref \cite{R-7} the authors applied the van Hove limit, see below, to all relevant operators.  This gave explicit expressions for, e.g., the electrical conductivity and the   scattering-independent part of the response  function. However, the theory developed is valid only  for homogeneous systems and the  scattering-dependent part of the response  function   was not evaluated. In this paper we develop a QLRT for  inhomogeneous  systems and investigate the influence of this randomness term (or scattering) on   plasmons in graphene and the  two-dimensional electron gas (2DEG). 

The paper is organized as follows.  In Sec. II we present some general QLRT expressions  and obtain  the DDRF as a sum of two terms, one of which is the usual term that is  independent of scattering and one that depends on it. In Sec. III we  investigate plasmons in graphene  and in Sec. IV plasmons in the 2DEG. In  Sec. IV we also  contrast its results with those for graphene. A summary follows in Sec. V and some important results of Sec. II are detailed  in the appendix.

\section{Formalism }

We consider a system whose Hamiltonian is 
\cite{R-1},
\begin{equation}\label{e:1}
H=H_0-AF(t)+\lambda V,
\end{equation}
where $H_0$ is the system's  Hamiltonian in the absence of external stimuli 
and randomness. The external time-dependent probe  $F(t)$ couples to the operator  $A$ and   randomness or scattering is represented by $\lambda V$;  this could be, for instance,  electron-impurity or electron-phonon interaction, and $\lambda$  is  strength of this interaction.  Be in the linear-response regime is expressed by the inequality 
\begin{eqnarray}\label{e:2}
\langle AF(t)\rangle  &, & \langle \lambda V\rangle \ll \langle H_0\rangle.
\end{eqnarray}
Many  optical and transport properties  can  be obtained from the knowledge 
of the density-density response function (DDRF) and we focus on obtaining it. In this case the relevant operator $A$ in Eq. (\ref{e:1}) is the density operator $\rho$. Furthermore, the external probe is the potential  $V(\vec{r},t)$. The second term in Eq. (\ref{e:1}) becomes 
\begin{equation}\label{e:3}
AF(t)\equiv\int V\left(\vec{r},t\right)\rho(\vec{r})d\vec{r}.
\end{equation}
The general expression for the DDRF is \cite{R-9} 
\begin{equation}\label{e:4}
\chi^0(\vec{r},t;\vec{r^{'}},t')=-\frac{i}{\hbar}\Theta(t-t')\Big\langle \left[\rho(\vec{r},t),\rho(\vec{r^{'}},t')\right]\Big \rangle_0 ,
\end{equation}
where $\langle ... \rangle_0$ denotes an average over a  thermal equilibrium ensemble. $\chi^0(\vec{r},t;\vec{r^{'}},t')$ is proportional to the probability of finding an electron at position $\vec{r'}$ and time $t'$ knowing its position 
$\vec{r}$ at time $t$. The density operator is 
\begin{equation}\label{e:5}
\rho(\vec{r},t)=\psi^{\dagger}(\vec{r},t)\psi(\vec{r},t)=\sum_{ij}c_i^{\dagger}(t)c_j(t)\phi_i^{*}(\vec{r})\phi_j(\vec{r}),   
\end{equation}
with $\phi$ being the single particle wave function. In the absence of the randomness term, $\lambda V$, the time evolution of an operator in the Heisenberg picture is  
\begin{equation}\label{e:6}
c(t)=e^{iH_0t/\hbar}ce^{-iH_0t/\hbar}.
\end{equation}
We now rename   $\chi$ as $\chi_{non}$ to distinguish it from the part that depends on scattering, see below. Employing Eqs. (\ref{e:5}) and (\ref{e:6}), the DDRF in the absence of randomness in the frequency domain is given by \cite{R-10}
\begin{equation}\label{e:7}
\chi_{non}^0(\vec{r},\vec{r^{'}},\omega)=\lim_{\nu \rightarrow 0}\sum_{i,j}\Delta_{ij}\phi_i^{*}(\vec{r})\phi_j(\vec{r})\phi_j^{*}(\vec{r^{'}})\phi_i(\vec{r^{'}}),  
\end{equation}
where 
\begin{equation}
\Delta_{ij}\equiv\frac{f_i-f_j}{E_i-E_j+\hbar(\omega+i\nu)}, 
\end{equation}
and $f$ stands for the Fermi-Dirac distribution function. The van Hove limit, described in Refs. \cite{R-1}, is given by 
\begin{eqnarray}\label{e:8}
\lambda \rightarrow  0 &,  t/\tau_t \rightarrow \infty,& \lambda^2t= finite,
\end{eqnarray}
where  $\tau_t$ is the transition time. This limit alters the time evolution of  an operator \cite{R-1} in the manner 
\begin{equation}\label{e:9}
c_l(t)=e^{-\Lambda_l t}c_l+e^{iH_0t/\hbar}c_le^{-iH_0t/\hbar},
\end{equation}
where $\Lambda$ is a super-operator  defined by 
\begin{equation}\label{e:10}
\Lambda \hat{c}\equiv \sum_{i,j}|i\rangle\langle i|\left[W_{ji}\langle j|\hat{c}|j\rangle-W_{ij}\langle i|\hat{c}|i\rangle\right];
\end{equation}
 the transition rate $W_{ij}$ given by Fermi's golden rule 
\begin{equation}\label{e:11}
W_{ij}=\left(2\pi\lambda^2/\hbar\right)|\langle i|v|j\rangle|^2\delta\left(E_i-E_j\right).
\end{equation} 
Note that in  Ref. \cite{R-7} the calculation has been done in a representation in which all operators have, in principle, a diagonal and a nondiagonal part. Here we exploit the approach used in  Refs. \cite{R-6} and \cite{R-9}. It is worth mentioning that Eq. \eqref{e:9}  satisfies the equation of motion, $i\hbar\partial_t{c_l(t)}=[c_l(t),H_0+\lambda V]$  in the van Hove limit. Applying Eqs. (\ref{e:5}) and (\ref{e:9}) the DDRF takes the form 
\begin{equation}\label{e:12}
\chi^0(\vec{r},\vec{r^{'}},\omega)=\chi_{im}^0(\vec{r},\vec{r^{'}},\omega)+\chi^0_{non}(\vec{r},\vec{r^{'}},\omega),
\end{equation}
with 
\begin{equation}\label{e:13}
\chi_{im}^0(\vec{r},\vec{r^{'}},\omega)=\sum_{i,j}M_{ij}\phi_i^{*}(\vec{r})\phi_j(\vec{r})\phi_j^{*}(\vec{r^{'}})\phi_i(\vec{r^{'}}), 
\end{equation}
and  
\begin{equation}\label{e:14-1}
M_{ij}\equiv\Big\langle\frac{1}{\hbar\left(\omega+i\left(\Lambda_{i}+\Lambda_{j}\right)\right)}\Big\rangle_b (f_i-f_j); 
\end{equation}
here $\langle ..\rangle_b$ denotes an average over the boson bath states. The details of the calculation are given in appendix A.

 \section{Plasmons in graphene} 
From  the  DDRF many properties such as plasmons, 
reflection and transmission amplitudes  can be evaluated \cite{R-6,R-9,R-10,R-11}. Below we investigate  graphene plasmons in the  random-phase approximation (RPA) \cite{R-12}. For low energies graphene's energy spectrum  is given by \cite{R-13}, 
\begin{equation}\label{e:14}
E_\zeta(\vec{k})=\zeta\hbar v_F k,  
\end{equation}
where $\zeta=-1(+1)$ indicates the valence (conduction) band. The corresponding  single-particle wave function is 
\begin{equation}\label{e:15}
\psi_{\zeta \mu s \vec{k}}(\vec{r})=\frac{1}{\sqrt{2A}}\begin{pmatrix} 
e^{-i\mu \theta(\vec{k})} \\
\zeta
\end{pmatrix}e^{i\vec{k}.\vec{r}}X_s  
\end{equation}
with $\theta(\vec{k})=\tan^{-1}(k_y/k_x)$,  
 $X_s$    the spin-dependent part of the wave function, and $\mu=1 (-1)$ the valley index for $K$ ($K'$).  In the long-wavelength limit,  at zero temperature, and for single-band (SB) transition, also known as intraband transition in the literature, $\chi^{0, SB}_{non}$ for an electron-doped system is given by \cite{R-13}
\begin{equation}\label{e:16}
\chi^{0, SB}_{non}(q',\omega')=\frac{k_F}{\pi\hbar v_F}\frac{q'^2}{\omega'^2}.
\end{equation}
As for $\chi^{0, SB}_{im}$, by employing Eqs (\ref{e:14}), (\ref{e:15}), and (\ref{e:13}) we obtain%
\begin{equation}\label{e:17}
\chi^{0, SB}_{im}(q',\omega')=\frac{k_F}{2\pi\hbar v_F}\, C(\omega',\gamma')
\left(1-\delta_{\gamma',0}\right),
\end{equation}
 where we used the dimensionless parameters $q'$, $\omega'$, and $\gamma'$ $\left(q'\equiv q/k_F,\omega'\equiv \hbar\omega/E_F, \gamma'\equiv \hbar\gamma/E_F \right)$  to simplify all expressions. 
  $C(\omega',\gamma')$ is given below. Although $\langle \Lambda \rangle_b$ is a function of momentum, for simplicity in the derivation of Eq. \eqref{e:18} it  has been replaced  by $\tau/2$ where $\tau$ is the relaxation time and $\gamma=1/\tau$; this is valid only for elastic scattering. For  two-band (TB) transitions, known  as interband transitions, $\chi^{0, TB}_{non}$ \cite{R-13} is given by
\begin{equation}\label{e:18}
\chi^{0, TB}_{non}(q',\omega')=\frac{k_F}{\pi\hbar v_F}\frac{q'^2}{2\omega'} \Big[A(\omega') -i\frac{\pi}{2}\,\Theta(\omega'-2)\Big]
\end{equation}
and  $\chi^{0, TB}_{im}$ by 
\begin{equation}\label{e:19}
\chi^{0,  TB}_{im}(q',\omega')=\frac{k_F}{\pi\hbar v_F}\,C(\omega',\gamma')
\left(1-\delta_{\gamma',0}\right).
\end{equation}
where 
\begin{equation}\label{e:20}
\hspace*{-0.3cm}A(\omega')=\frac{2}{\omega'}+\frac{1}{2}\ln\big|\frac{2-\omega'}{2+\omega'}\big|,\,\,\, C(\omega',\gamma')=\frac{\omega'-i\gamma'}{\omega'^2+\gamma'^2}.
\end{equation}
The DDRF in momentum and energy space, $\chi(q',\omega')$, characterizes the probability to find an electron which its final and initial states differing in momentum and energy by $q'$ and $\omega'$, respectively. In other words, it describes the probability of an electron excitation. 

\begin{figure}[b]	
	\includegraphics[height=4cm, width=6cm]{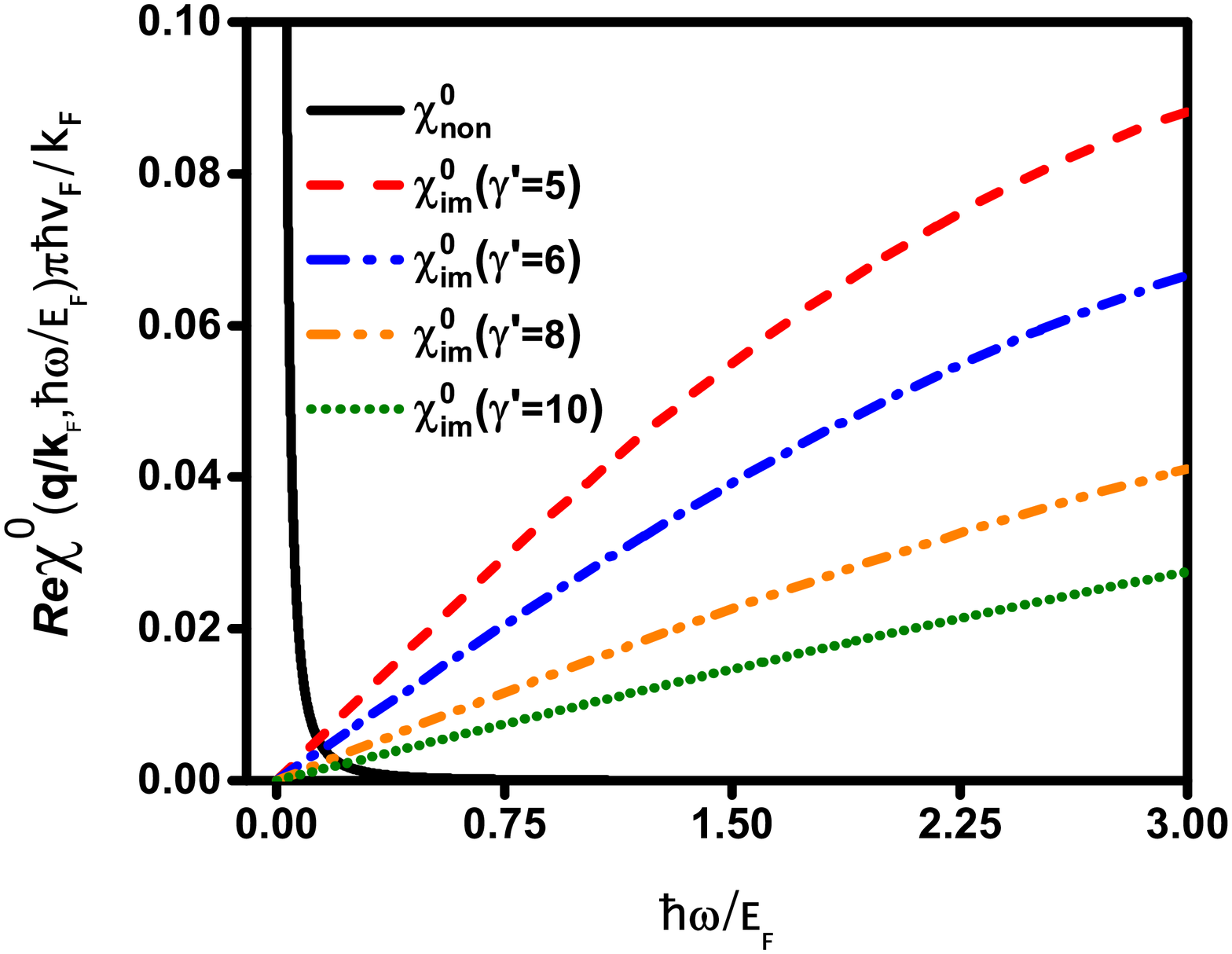}	
	\caption{\small{Real parts of  $\chi_{non}$ and $\chi_{im}$ versus $\omega'$ where $\chi_{im}$ is displayed for several $\gamma'$.}}
	\label{fig :0}	
\end{figure}

We show $\chi_{non}$ and $\chi_{im}$ as functions of $\omega'$ in Fig. \ref{fig :0}.  Actually, Fig. \ref{fig :0} shows that the magnitude of the real part of Eq. \eqref{e:19}  ,$\mathfrak{Re}\chi_{im}$  dominates  for almost all $\omega'$ except for very small $\omega'$ for which the magnitude of $\mathfrak{Re}\chi_{non}$ is larger that $\mathfrak{Re}\chi_{im}$.  Therefore, apart for very small $\omega'$ one can obtain  all properties of graphene, related to $\chi$, from $\chi_{im}$ that has not been considered so far. In addition, as seen in Fig. \ref{fig :0}, for fixed $\omega'$ increasing $\gamma'$ makes  $\chi_{im}$ weaker since increasing $\gamma'$ leads to shorter scattering time and  length. Therefore, the probability for  an electron to reach the final  desired momentum  is reduced by strengthening the interaction with impurity. 

\begin{figure}[ht]
	\vspace*{0.5cm}
	\begin{center}		
		\includegraphics[height=4cm, width=4cm]{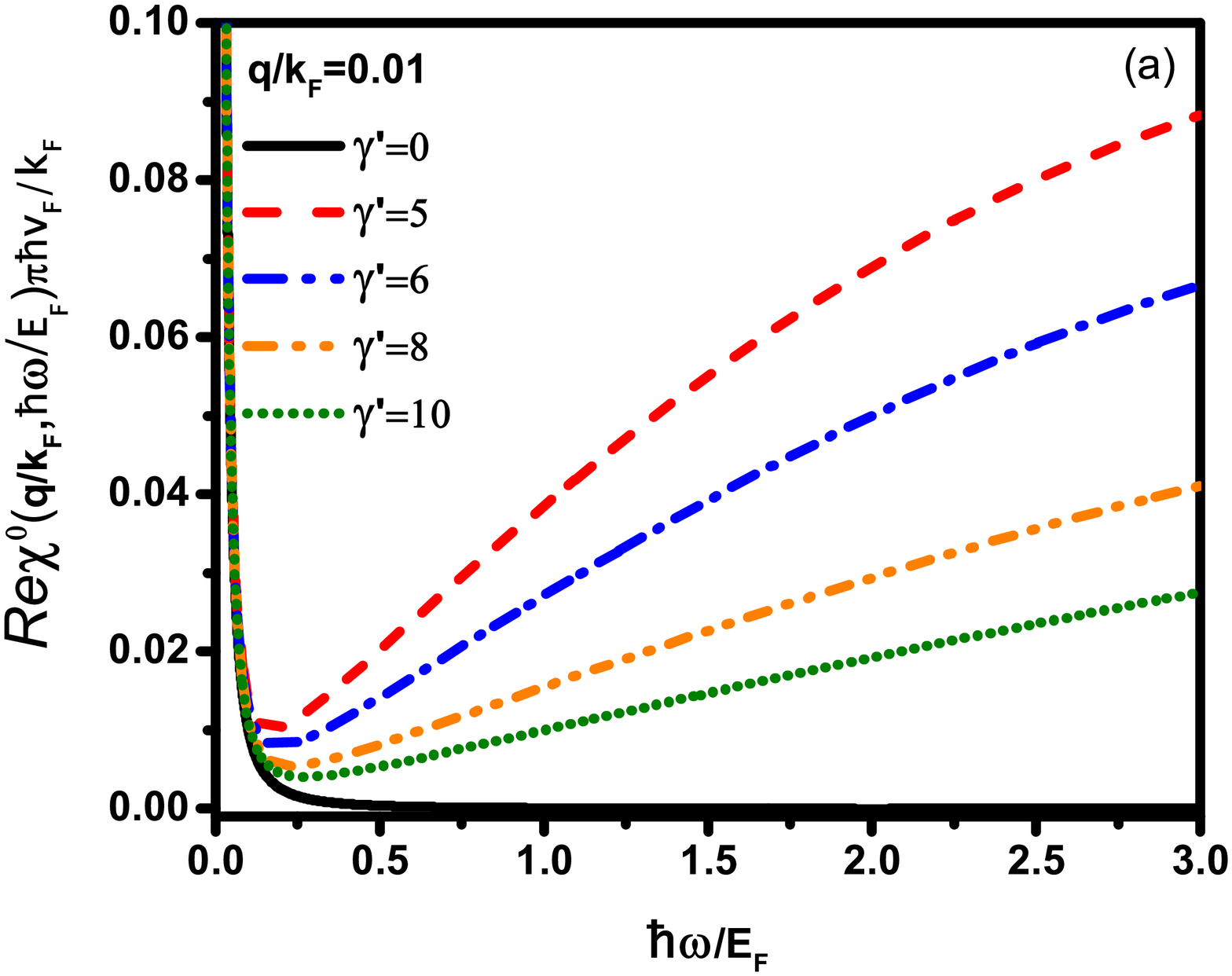}\quad
		\includegraphics[height=4cm, width=4cm]{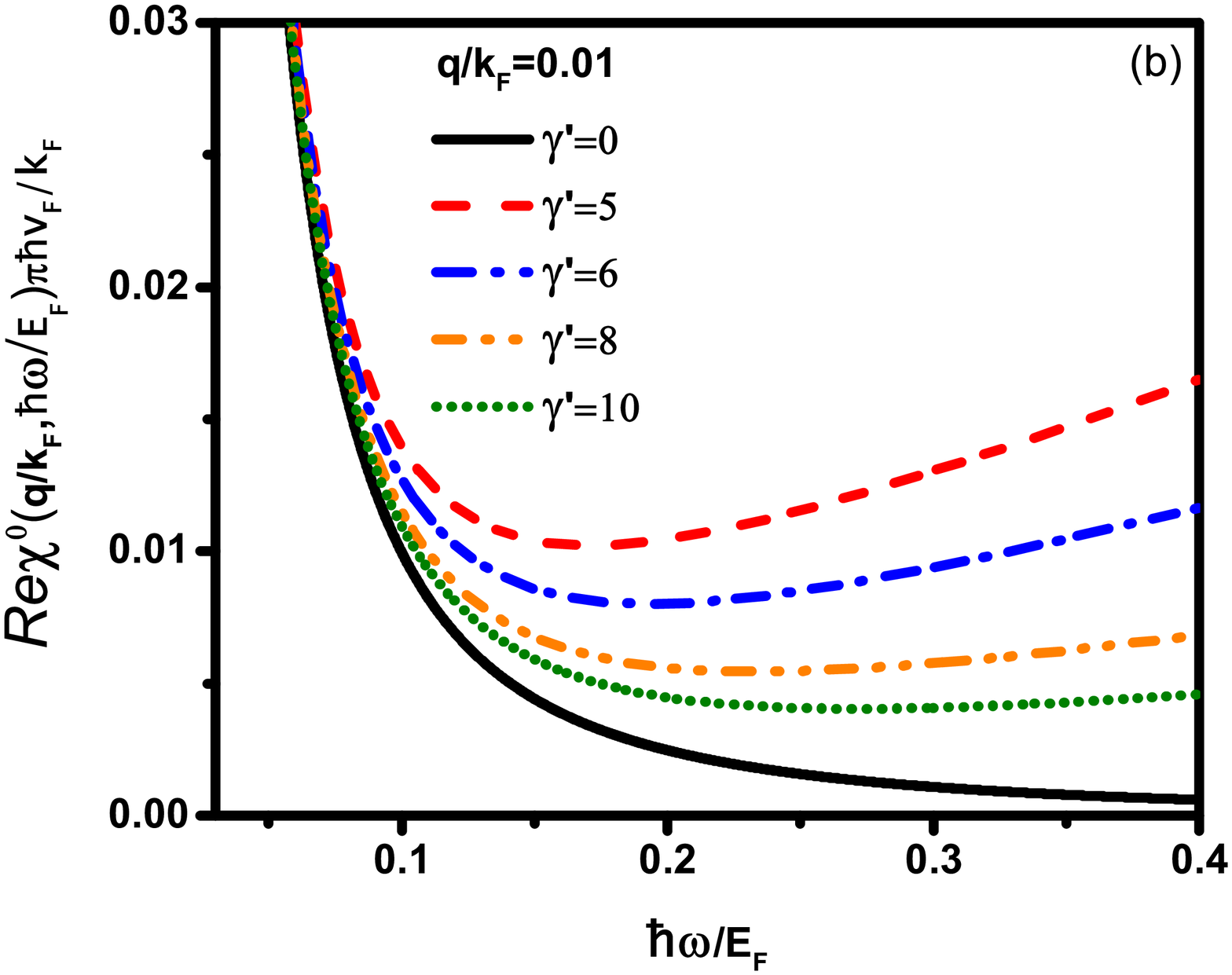}
	\end{center}	
	\caption{\small{(a)  Real part of the TB DDRF ($\mathfrak{Re}\chi$) for several values of $\gamma'$. (b)  A part of (a) for $\omega'\leq0.4$. }}
	\label{fig :1 }
\end{figure}

The real part of the total  TB  DDRF ($\mathfrak{Re}\chi$), containing both $\chi_{non}$ and $\chi_{im}$, is shown in Fig. \ref{fig :1 } (a) versus $\omega'$  for  a typical value of $q'$  in the long-wavelength limit. The solid black curve represents the TB DDRF without inclusion of scattering whereas the coloured curves are for several different $\gamma'$. To make clearer its dependence on  $\omega'$ in  Fig. \ref{fig :1 } (b) we blow up the part of  Fig. \ref{fig :1 } (a) for $\omega'\leq 0.4$. As seen,  the TB DDRF for small $\omega'$ decreases dramatically because in this energy range  the principal contribution to it emanates from  $\chi_{non}$ as we mentioned in the justification of Fig.\ref{fig :0}.

\begin{figure}
	\begin{center}		
		\includegraphics[height=4cm, width=4cm]{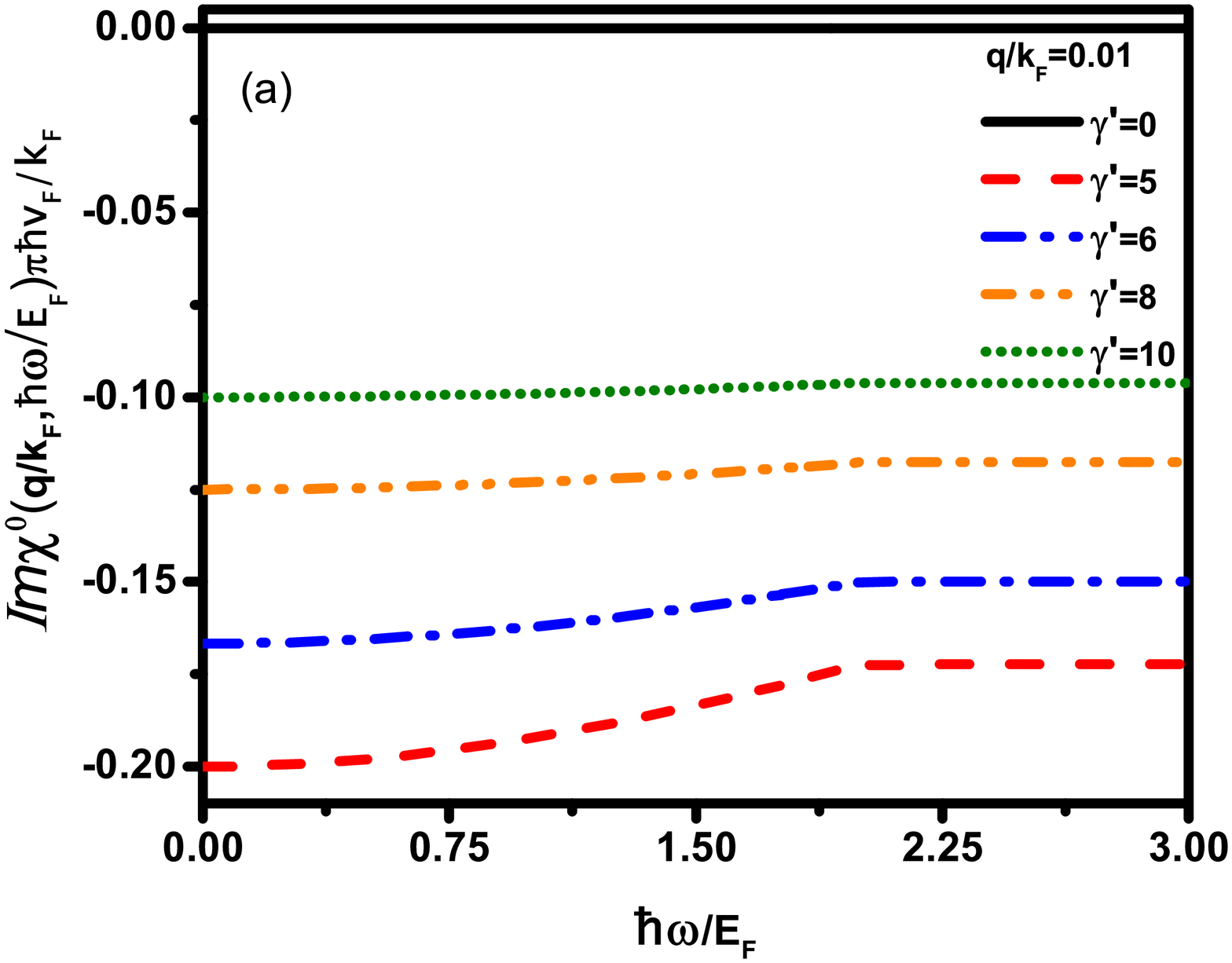}\quad	
		\includegraphics[height=4cm, width=4cm]{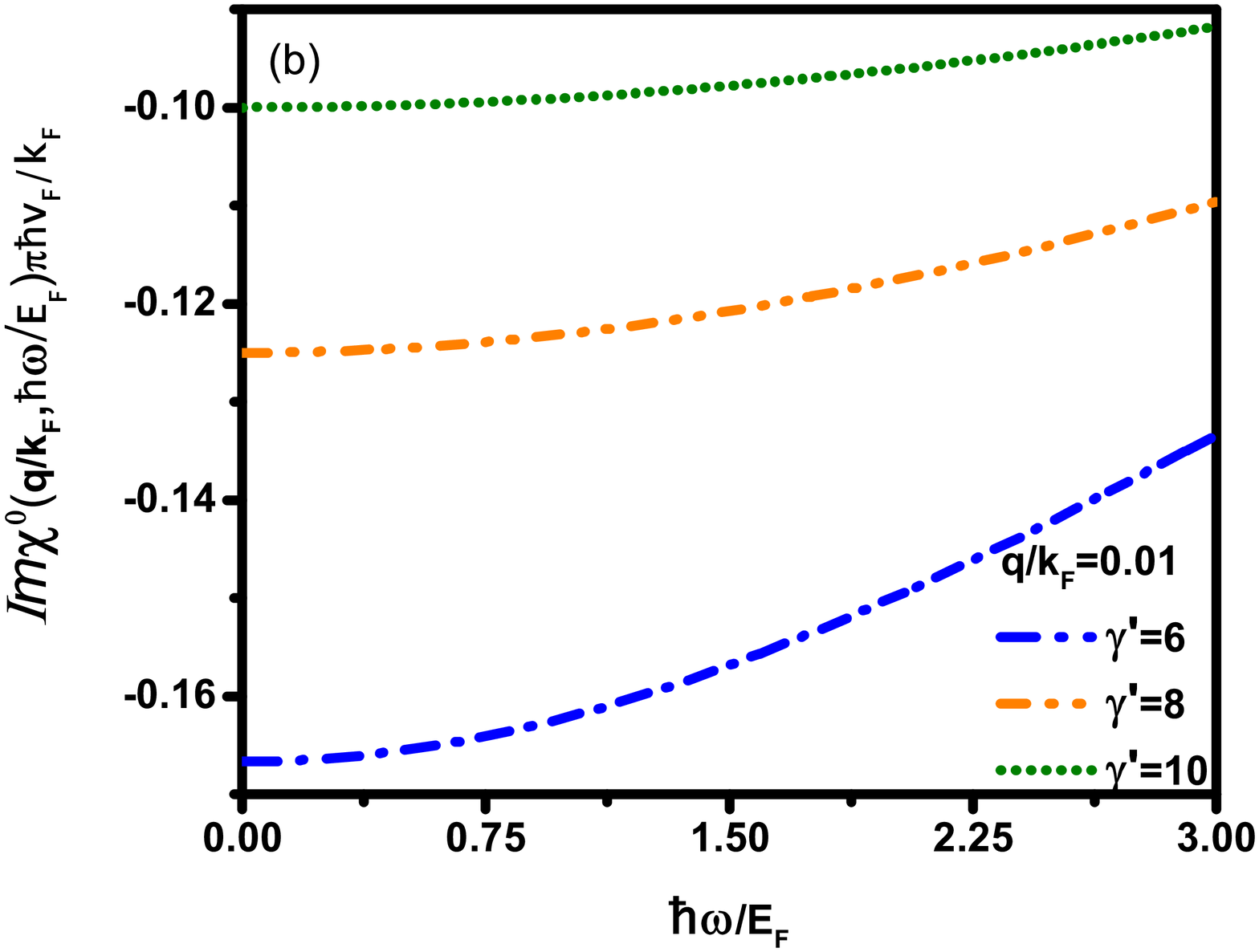}
		\end{center}
		\caption{\small{(a)  Imaginary part of the TB DDRF, $\mathfrak{Im}\chi$, for $q'=0.01$ and several 	$\gamma'$. (b) A segment of $\mathfrak{Im}\chi$ for three $\gamma'$.  }}
		\label{fig :2 }	
\end{figure}

The imaginary part of the TB DDRF ($\mathfrak{Im}\chi$), versus $\omega'$, is shown in Fig. \ref{fig :2 } (a)  for several $\gamma'$. To make more transparent  its dependence on $\omega'$, in Fig. \ref{fig :2 } (b)  we show it for three different $\gamma'$ on an expanded scale.  Notice though that this makes the cusps or maxima of Fig. 3(a) invisible.

\begin{figure}
		\includegraphics[height=4cm, width=4cm]{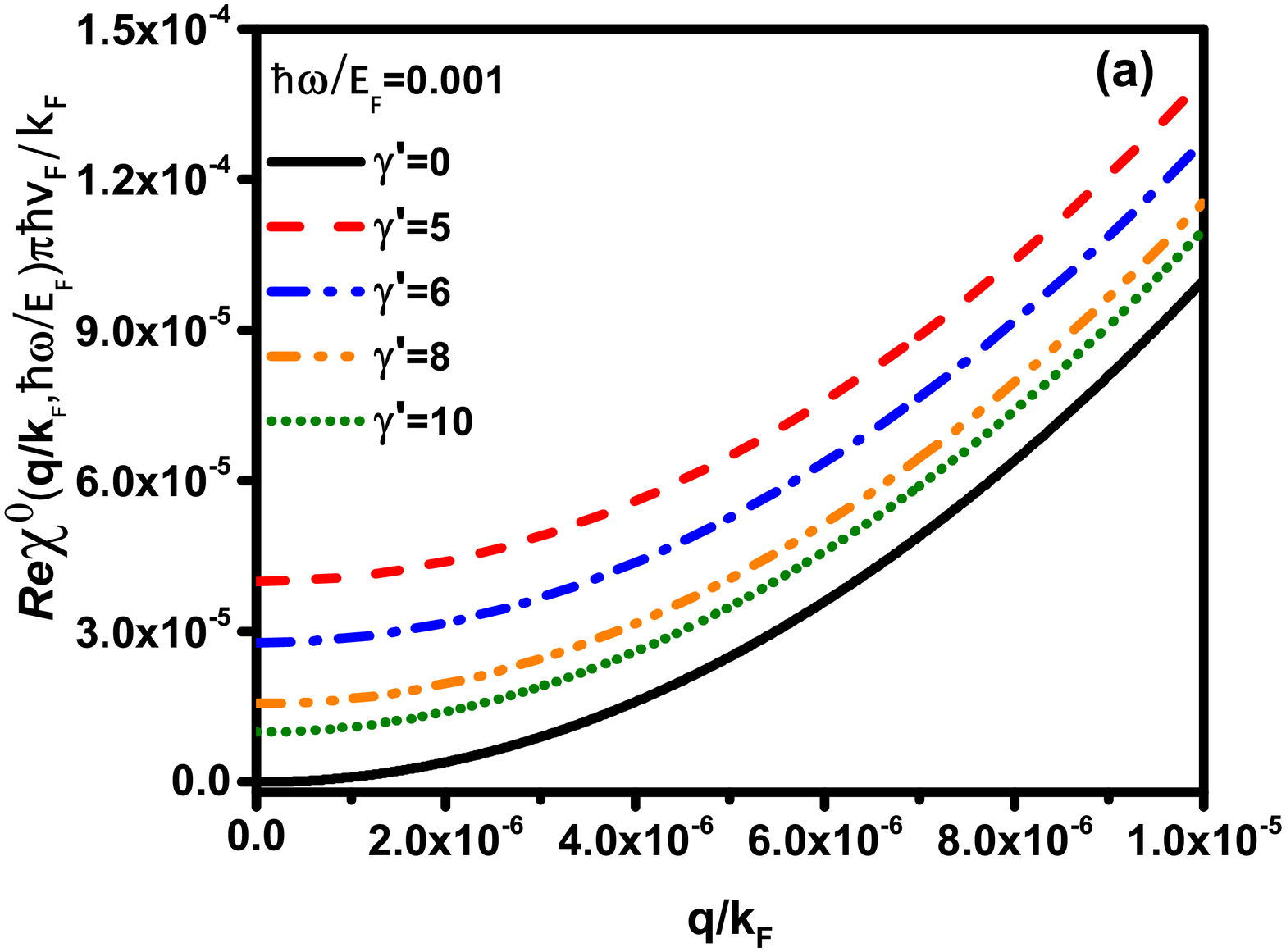}\quad
		\includegraphics[height=4cm, width=4cm]{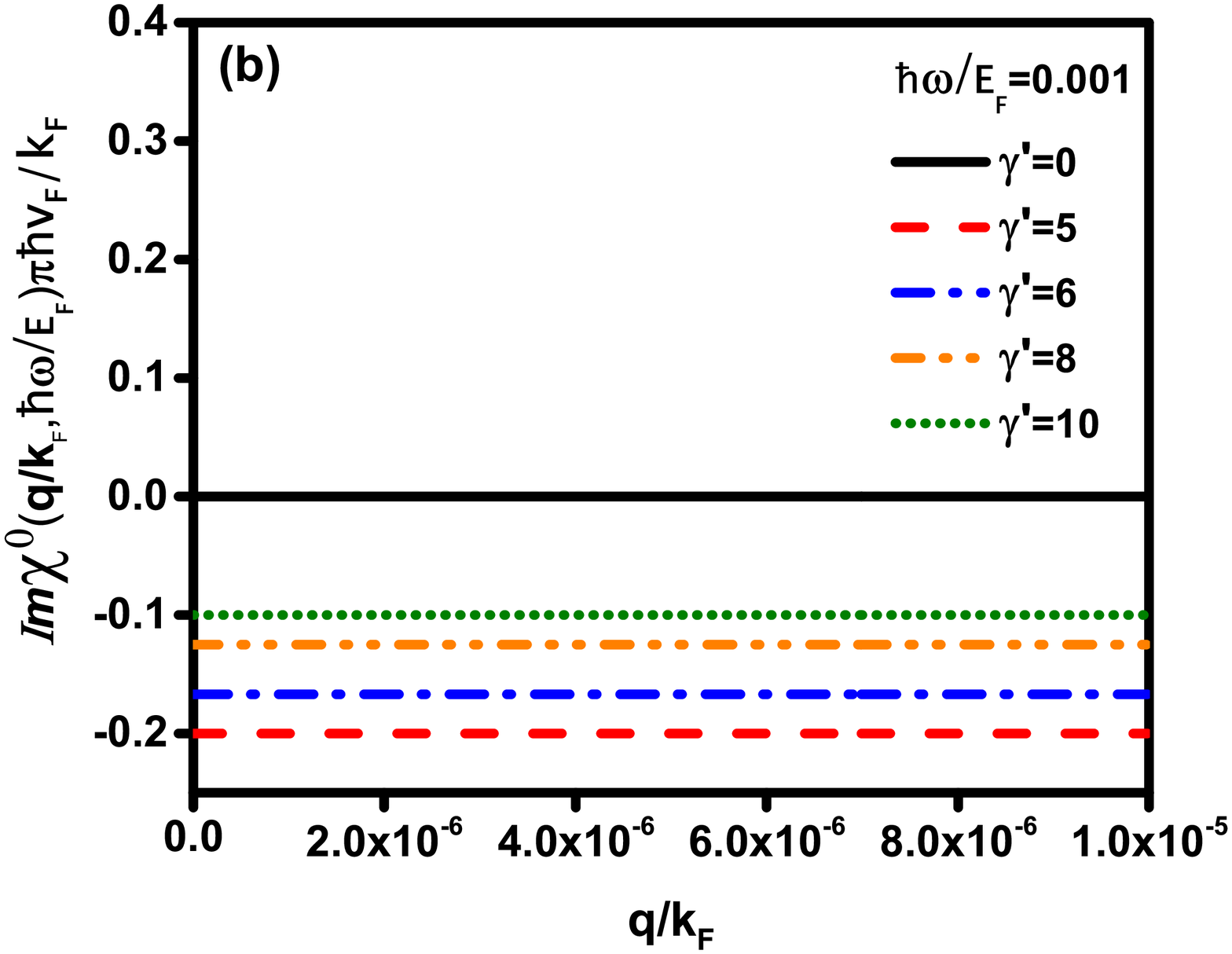}								
		\caption{\small{(a)  Real and (b) imaginary parts of the TB  DDRF versus $q'$ at $\omega'=0.001$ for  several different values of $\gamma'$} }
			\label{fig :3 }	
\end{figure}

Figures  \ref{fig :3 } (a) and (b) show the real and imaginary parts of the TB DDRF versus $q'$ for several different values of $\gamma'$ and a typical  frequency $\omega'=0.001$. As shown in \ref{fig :3 } (a) 
the  dependence of $\mathfrak{Re}\chi$ is approximately  parabolic because only $\chi_{non}$ has a term that contains $q'$ and contributes to the TB DDRF. In addition, it is clear that  for $q'$ fixed  $\mathfrak{Re}\chi$ increases as $\gamma'$  decreases. 

\begin{figure}[t]
		\includegraphics[height=4cm, width=4cm]{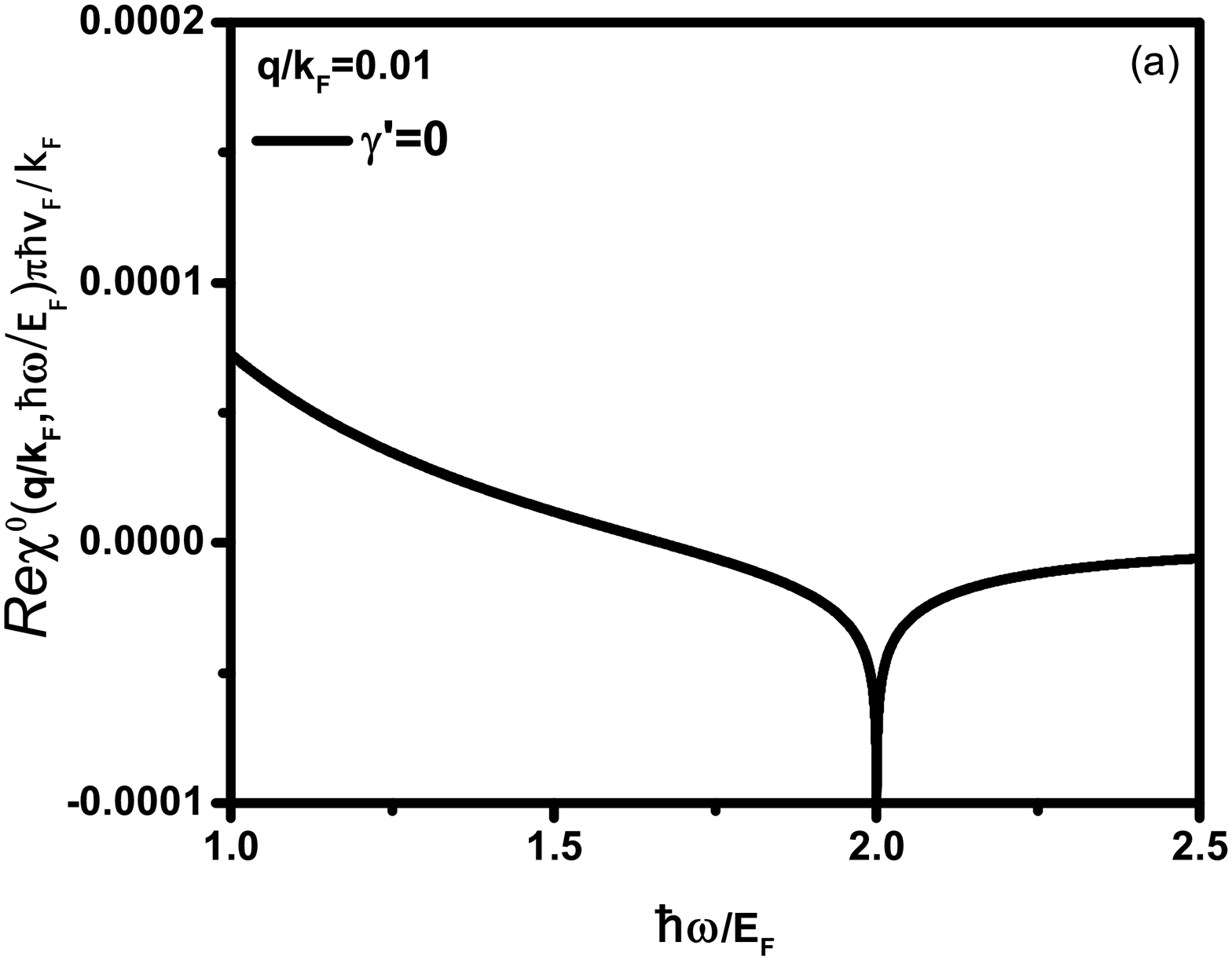}\quad	
		\includegraphics[height=3.9cm, width=4cm]{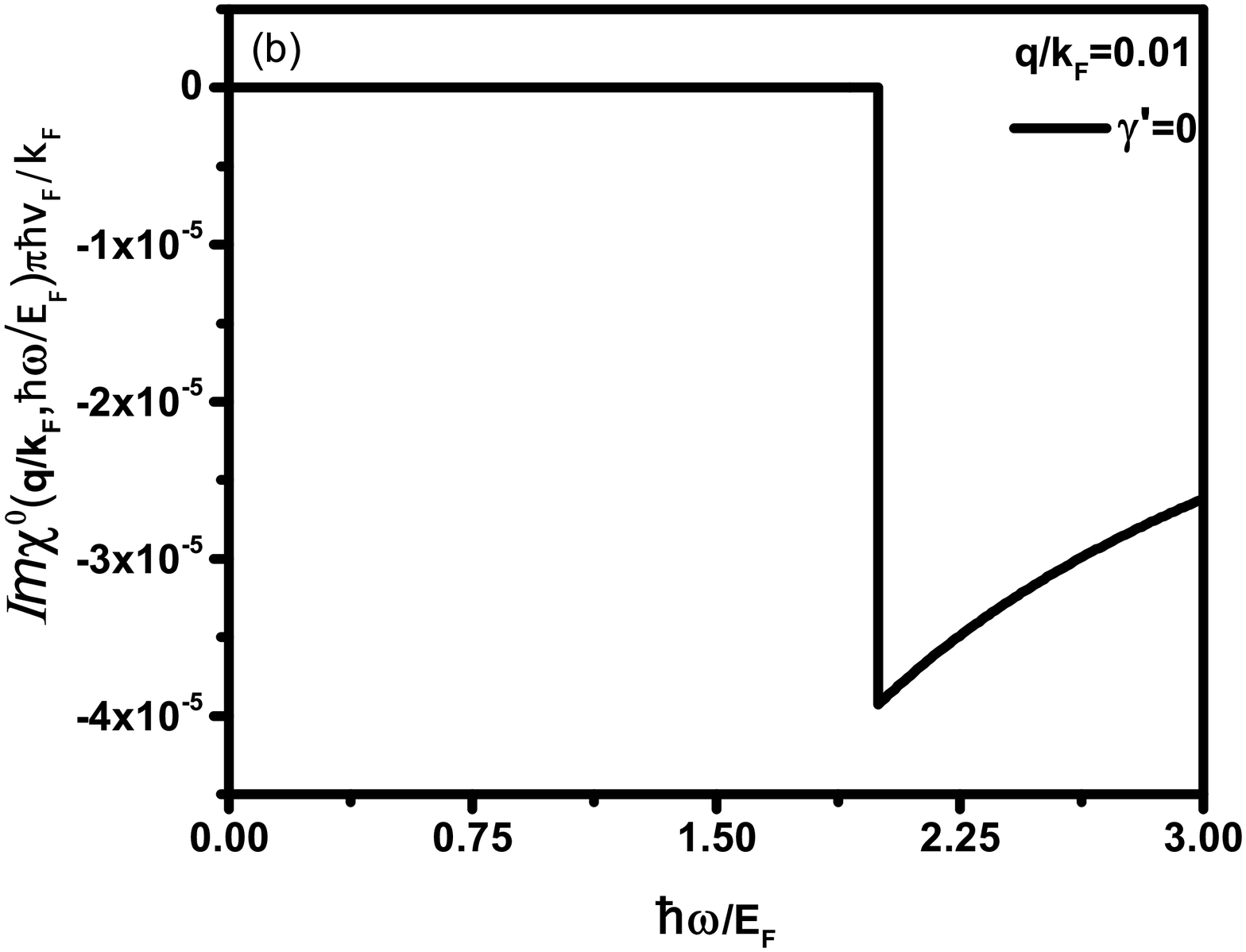}	
	\includegraphics[height=4cm, width=4cm]{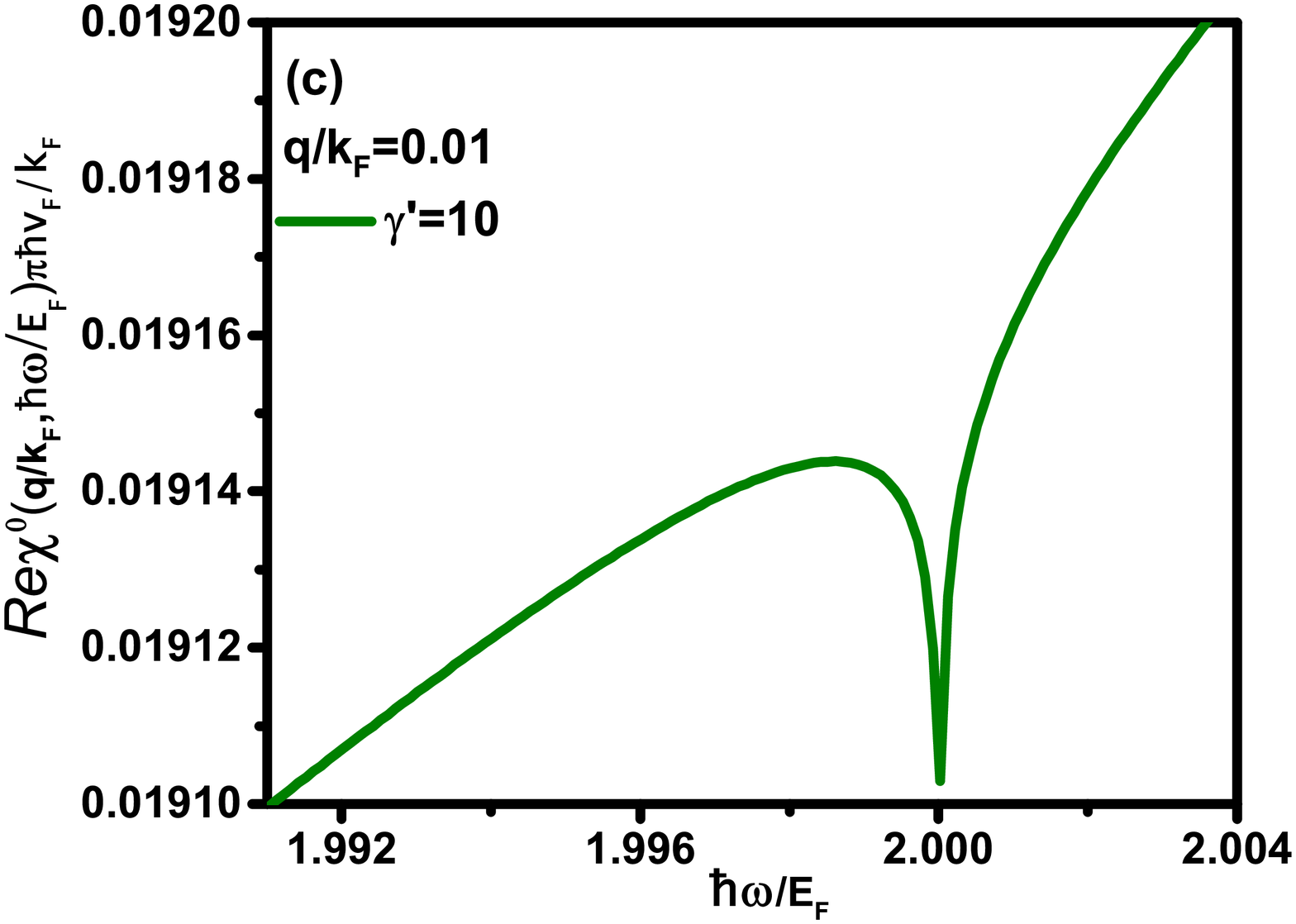}\quad	
	\includegraphics[height=3.9cm, width=4cm]{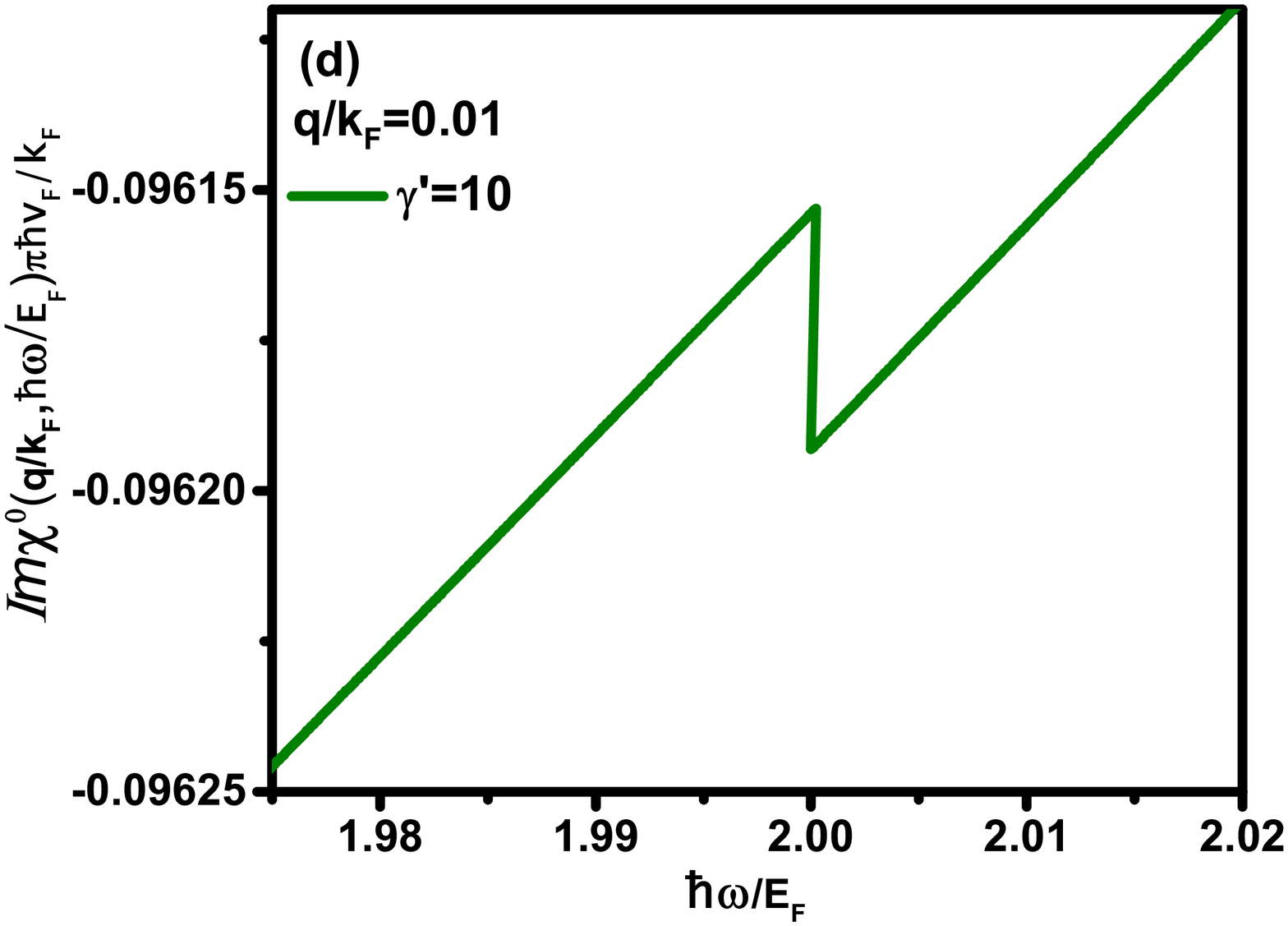}	
	\caption{\small{(a) Real and (b) imaginary parts of  the TB DDRF  for $\gamma'=0$ versus $\omega'$.  (c)- (d): as in (a) and (b)  for $\gamma'=10$. } }
	\label{fig :4 }	
\end{figure}

Since the logarithmic and  step function terms  alter the behaviour of  $\mathfrak{Re}\chi$ and $\mathfrak{Im}\chi$, respectively, in the vicinity of $\omega'=2$ , which is not clear in  Figs.  \ref{fig :1 } (a) and \ref{fig :2 } (a) due to the large difference between their values with and without scattering, in Fig \ref{fig :4 } we display them separately. The upper panels are for $\gamma'=0$ and the lower ones for $\gamma'=10$. Notice i) how including scattering, $\gamma'\neq 0$, strengthens the behaviour of the results without it near $\omega' =2$ and ii)  without scattering ($\gamma'=0$)  $\mathfrak{Im}\chi$, shown in Fig. \ref{fig :4 }(b), vanishes for $\omega' \leq 2$ and that there is no   dissipation in the system. In contrast, when scattering is included $\mathfrak{Im}\chi$ has approximately a constant slope. 

The  permittivity $\epsilon$ of a system has a wealth of information. For example, the zeros of  its real part determine its plasmon dispersion. One of the approximations that  takes into account the electron-electron interaction is the RPA. The Fourier transformed, in space and time, RPA  permittivity 
is given by    
\begin{equation}\label{e:21}
\epsilon^{RPA}(q,\omega)=1-V(q)\chi^{0}(q,\omega),
\end{equation}
where $V(q)$ is the 2D Fourier transform of the Coulomb potential. 
In   terms of $q'$  we rewrite $V(q)$ as 
\begin{equation}\label{e:22}
V(q')=\eta/q',
\end{equation}
with $\eta\equiv 8\alpha\pi^2\hbar c/ \epsilon_b k_F$ and $\epsilon_b$  the background permittivity. For SB transitions 
the plasmons  can be  derived by combining Eqs.\eqref{e:16},\eqref{e:17}, and \eqref{e:21}. The result is 
\begin{equation}\label{e:23}
1-\frac{\beta}{q'}\left[\frac{q'^2}{\omega'^2}+\frac{\omega'}{2\left(\omega'^2+\gamma'^2\right)}\right]=0.
\end{equation}
One solution of this quadratic equation is
\begin{equation}\label{e:24}
q'=\frac{\omega'^2}{2\beta}\left(1+\left[1-\frac{2\beta^2\left(1-\delta_{\gamma',0}\right)}{\omega'\left(\omega'^2+\gamma'^2\right)}\right]^{1/2}\right),
\end{equation}
with $\beta\equiv \eta k_F/\pi\hbar v_F$, gives the dispersion relation. Notice that Eq. (26) gives the well-known dispersion $\omega\propto q^{1/2}$ for $\gamma=0$. The other  solution, with $1+$ in Eq. (26) replaced by $1-$, is unphysical and therefore rejected.

\begin{figure}[t]
		\includegraphics[height=4cm, width=4cm]{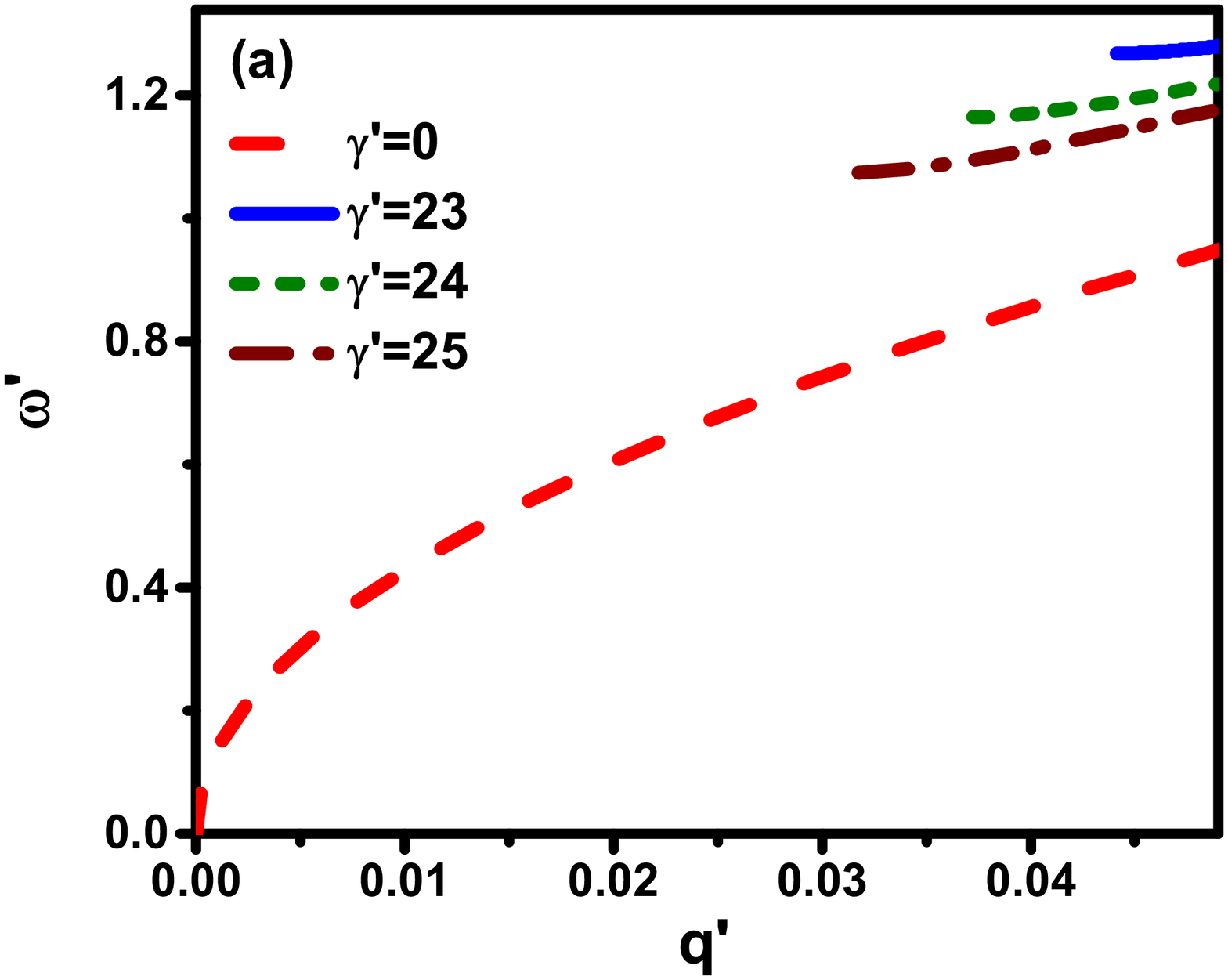}\quad	
		\includegraphics[height=4cm, width=4cm]{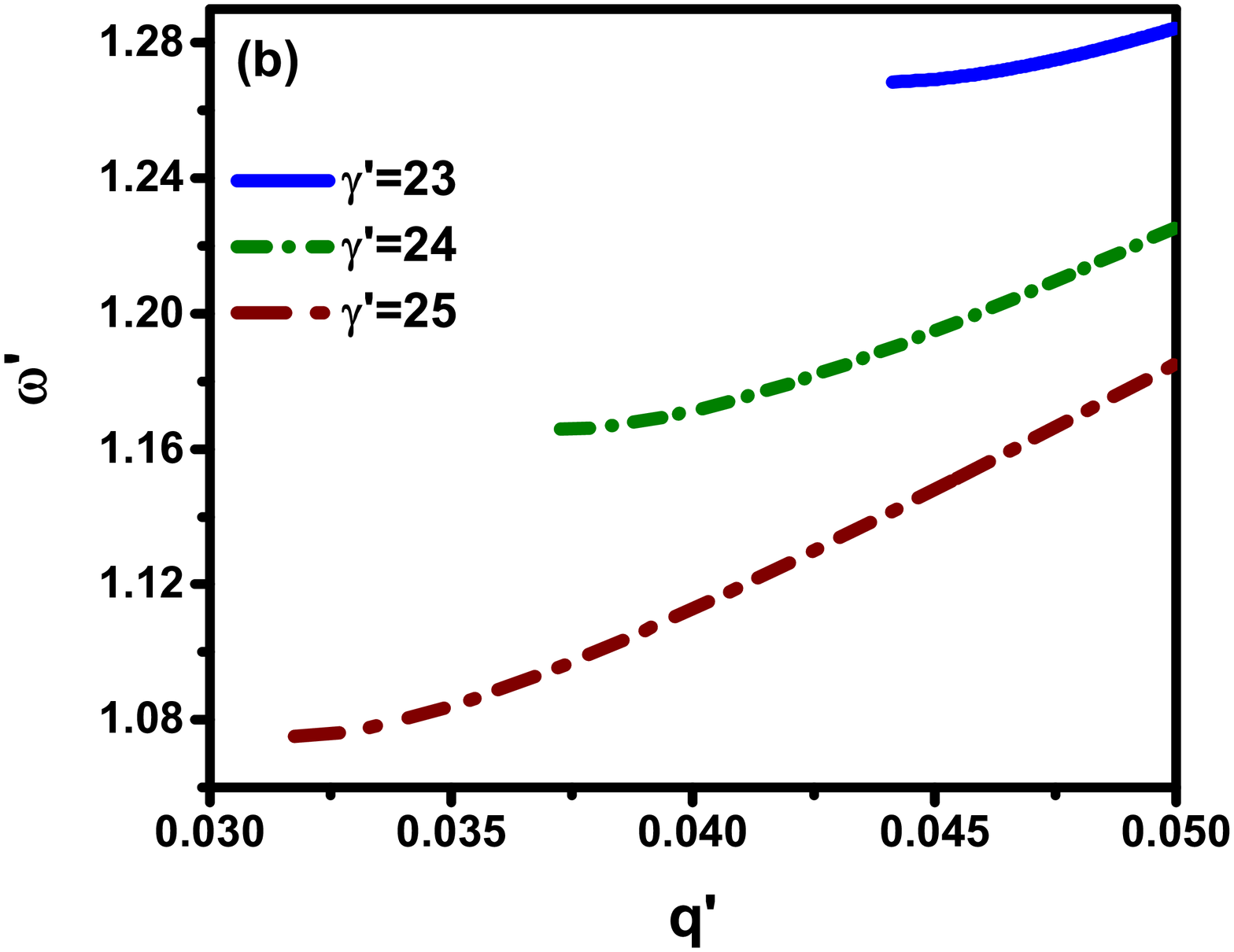}	
	\caption{{\small (a) SB dispersion relation for several values of $\gamma'$.  (b) The portion of (a) for $0.03\leq q^\prime\leq 0.05$.}}
	\label{fig:5}
\end{figure}

\begin{figure}[t]
		\includegraphics[height=4cm, width=4cm]{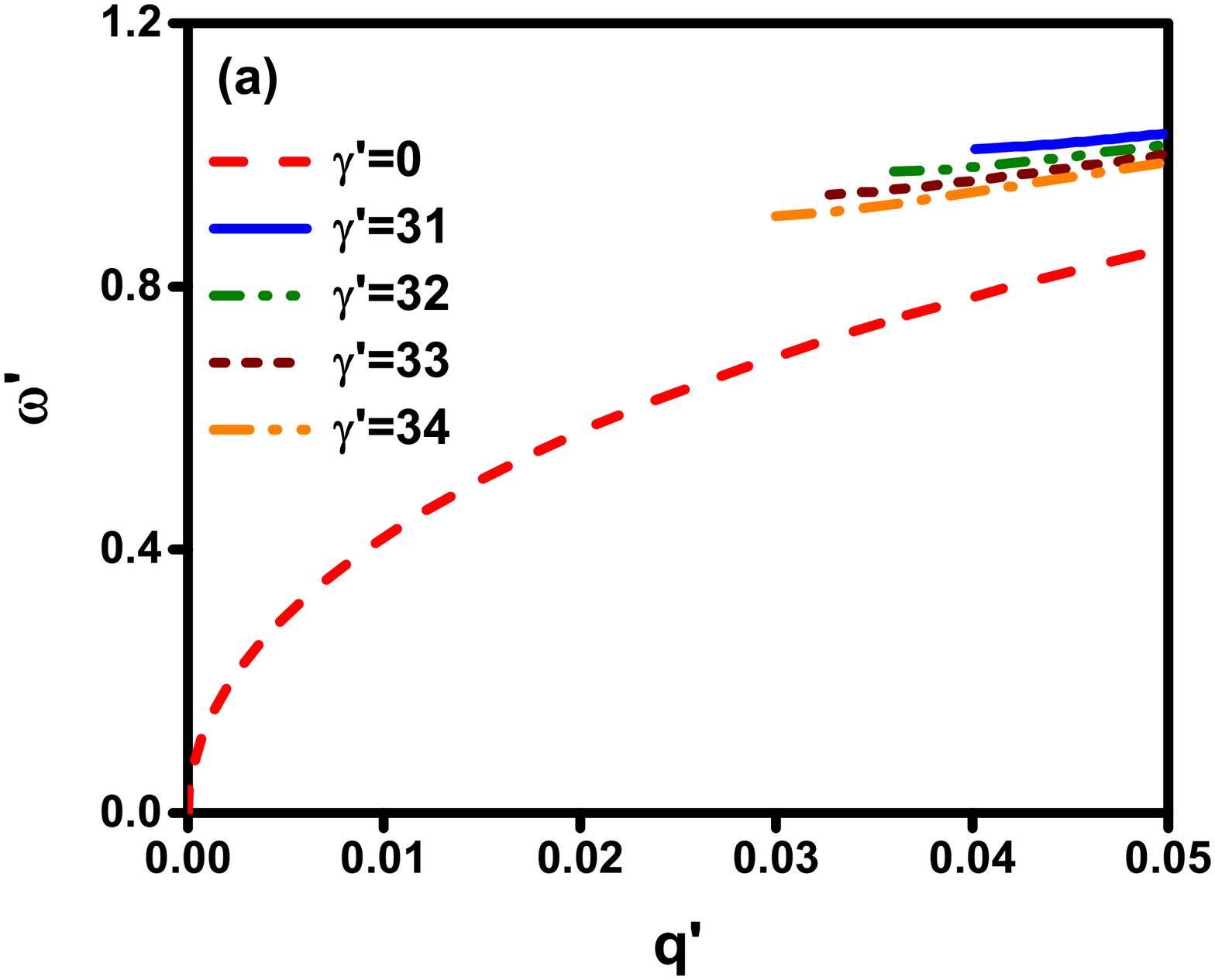}\quad	
		\includegraphics[height=3.9cm, width=4cm]{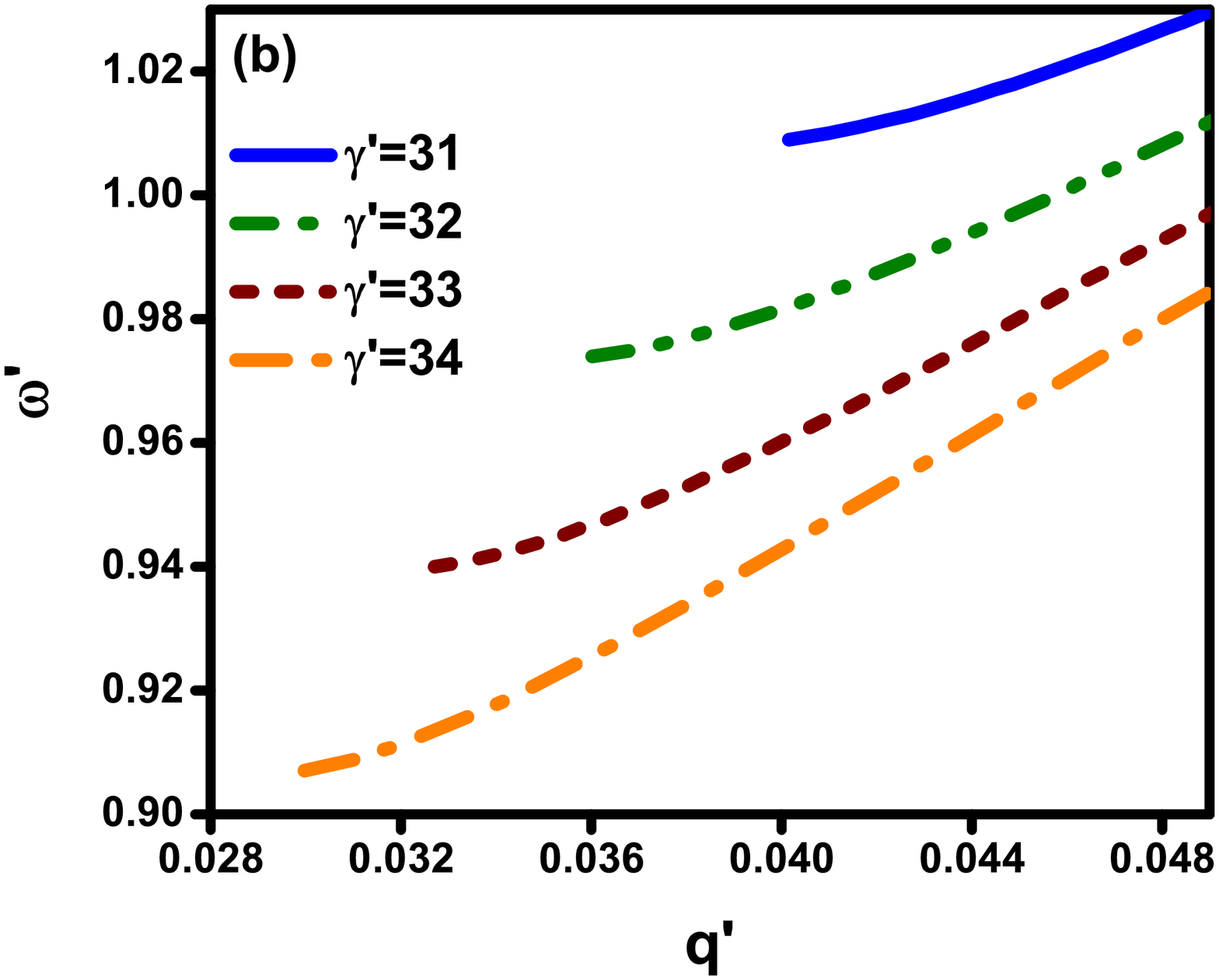}	
	\caption{\small{ (a) TB dispersion relation for several values of $\gamma'$.  (b) The portion of (a) for $0.028\leq q^\prime\leq 0.05$.}}	
	\label{fig:7}
\end{figure}

For TB transitions combining Eqs. \eqref{e:18},\eqref{e:19}, and \eqref{e:21} gives the plasmon spectrum as 

\begin{equation}\label{e:25-1}
\hspace*{-0.2cm}q'=\frac{\omega'}{\beta A(\omega')}\left[1+\left\{1-\frac{2\beta^2A\left(\omega'\right)\left(1-\delta_{\gamma',0}\right)}{\left(\omega'^2+\gamma'^2\right)}\right\}^{1/2}\right].
\end{equation}
Again the second solution for $q'$, with $1+$ in Eq. (27) replaced by $1-$, is unphysical and rejected.  In Figs. \ref{fig:5} (a) and \ref{fig:7} (a) we show the dispersion relations, resulting from Eqs. (26) and (27),  for several values of $\gamma'$.  To make these graphs more clear in Figs. \ref{fig:5} (b) and \ref{fig:7} (b) we show their  windows for $0.03\leq q^\prime\leq 0.05$ and different values of $\gamma'$, respectively.  In both cases the frequency  increases with $q'$ while the momentum plasmon range, i.e., the lower acceptable value of $q'$, decreases with increasing  $\gamma'$.  
Judging from the results as $\gamma$ increases in Figs. \ref{fig:5} and \ref{fig:7}, we see that  the SB plasmon frequency is larger than that the TB one, whereas  the SB plasmon momentum  range is a bit shorter than the TB one.  Note that the plasmon group velocity $d\omega/dq$ is approximately constant and independent of the  scattering strength $\gamma'$. 

A plasmon is a coherent collective excitation of the charge density  with all charges oscillating about their equilibrium positions. Scattering effects, such as electron-impurity or electron-phonon interaction, result in dissipation by single-particle excitations. In other words, a  single-particle excitation  competes with the collective  one:  if the mean-free path related to the single-particle excitation is of the order of the wavelength of the collective one,  there would be no plasmon. In Figs. \ref{fig:5} (b) and \ref{fig:7} (b) we can see that there is a critical plasmon momentum  below which there is no plasmon spectrum for a typical $\gamma'$.  This can be explained as follows.  The   mean-free path  decreases with increasing $\gamma'$.  By the uncertainty principle then its momentum increases with $\gamma'$ and so does the critical plasmon  momentum.  Physically, if the wavelength of the collective oscillation, the displacement from equilibrium, is smaller than the mean-free path,  the system supports plasmons.  

We also see, in both figures, that  for  fixed plasmon momentum the plasmon frequency increases with decreasing $\gamma'$. This  can be justified  as follows.  At fixed plasmon momentum   the coherent collective dipole momenta generate the plasmon electromagnetic field (EM) whose energy is determined  by the displacement from the equilibrium positions and the   number of available coherent dipole momenta. Higher impurity density, that is larger $\gamma'$,  increases the elastic scattering probability which  reduces the number of coherent dipole momenta. Therefore,  the plasmon EM field  resulting from them  will have  lower energy as $\gamma'$ increases.

\begin{figure}[hb]	
	\includegraphics[height=4cm, width=6cm]{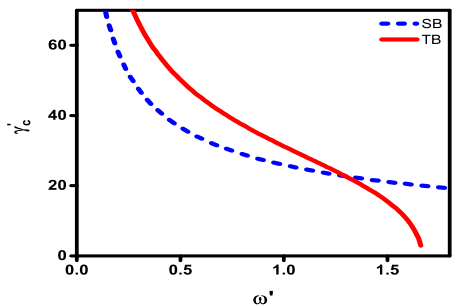}			
	\caption{\small{ SB and TB critical values   $\gamma'_c$ versus $\omega'$.} }	
	\label{fig:6-1}	
\end{figure}

As emphasized above, there are critical values  $\gamma'_c$   below which there are no  SB or TB  plasmons. To  find them we set the  factors $[...]^{1/2}$ and $\{...\}^{1/2}$ in Eqs. \eqref{e:24} and \eqref{e:25-1}, respectively,  equal to zero. This gives
\begin{align}
\gamma^{'SB}_c=\left[2\beta^2/\omega'-\omega'^2\right]^{1/2},
\end{align}
\vspace*{-0.5cm}
\begin{align}
\gamma^{'TB}_c=\left[2\beta^2A(\omega')-\omega'^2\right]^{1/2}.
\end{align}
\noindent  Figure \ref{fig:6-1} shows  $\gamma'_c$ versus the plasmon frequency $\omega'$ for SB and TB plasmons. In the former case the high value for  low frequencies decreases fast for $\omega'$ small but much more slowly for $\omega'>0.5$, while in the latter its value falls down very fast.

It is worth observing that setting  $[...]^{1/2}=0$ in Eq. (26) leads, for $\gamma\neq 0$ fixed, to a simple cubic equation for $\omega'$, $\omega'^3 +\gamma'^{2}\ \omega' -2\beta^2=0$. It's acceptable solution $\omega_c$ is given below  in Eq. (30). 	This then can be used to find analytically the lowest limit for $q_c=\omega_c^2/2\beta$, shown in Figs. 6 and 7, from Eq. (26).  The explicit results for $\gamma'\geq\gamma'_c$ are 
\begin{equation}
\omega_c=[\beta^2 +\sqrt{Z}]^{1/3}+[\beta^2 -\sqrt{Z}]^{1/3}, \,\, \,Z=\beta^{4}+\gamma'^6/27.
\end{equation}
Unfortunately, for TB transitions this is not possible due to the factor $A(\omega')$ in Eq. (27).

\begin{figure}[t]
		\includegraphics[height=4cm, width=4.2cm]{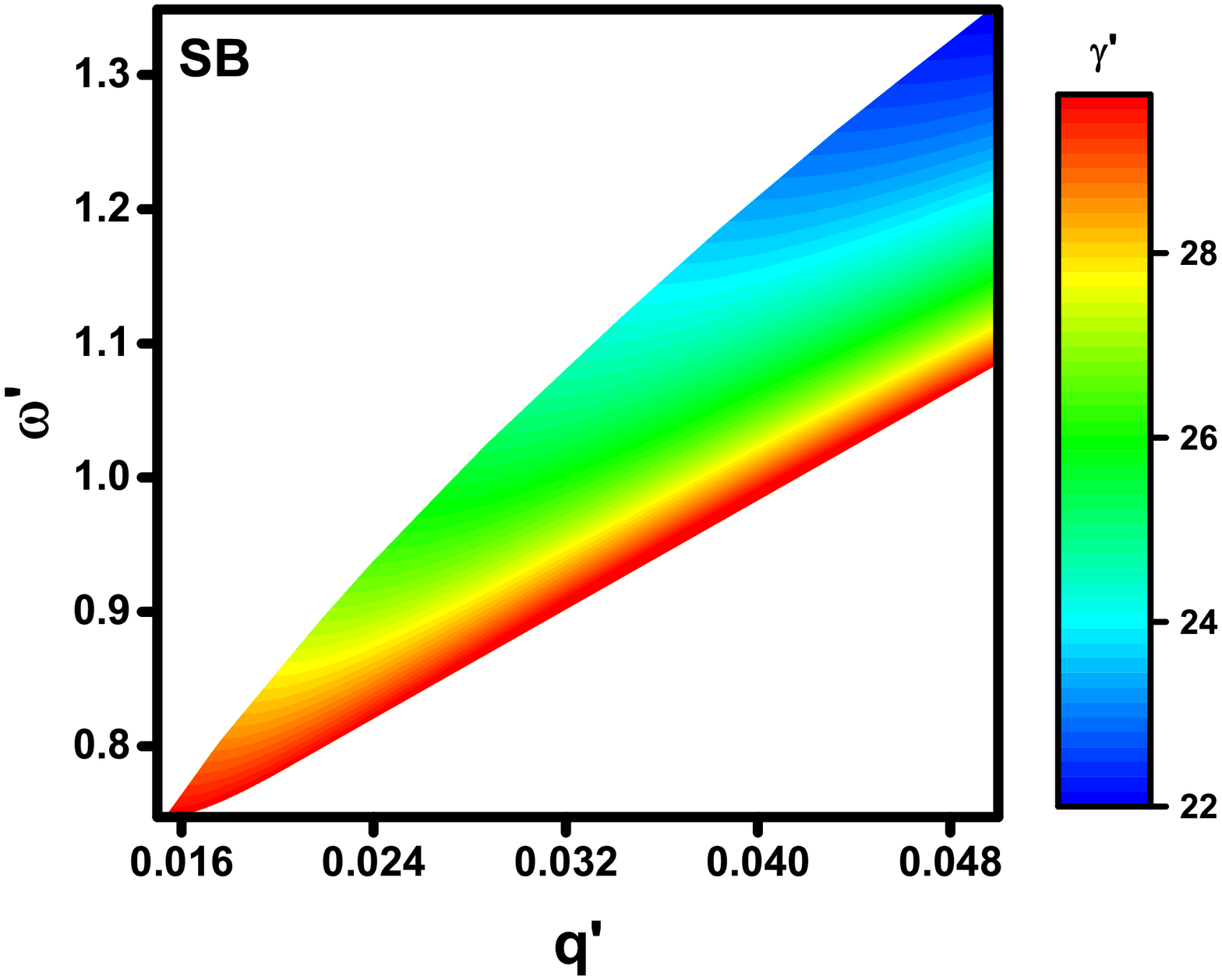}
		\includegraphics[height=4cm, width=4.2cm]{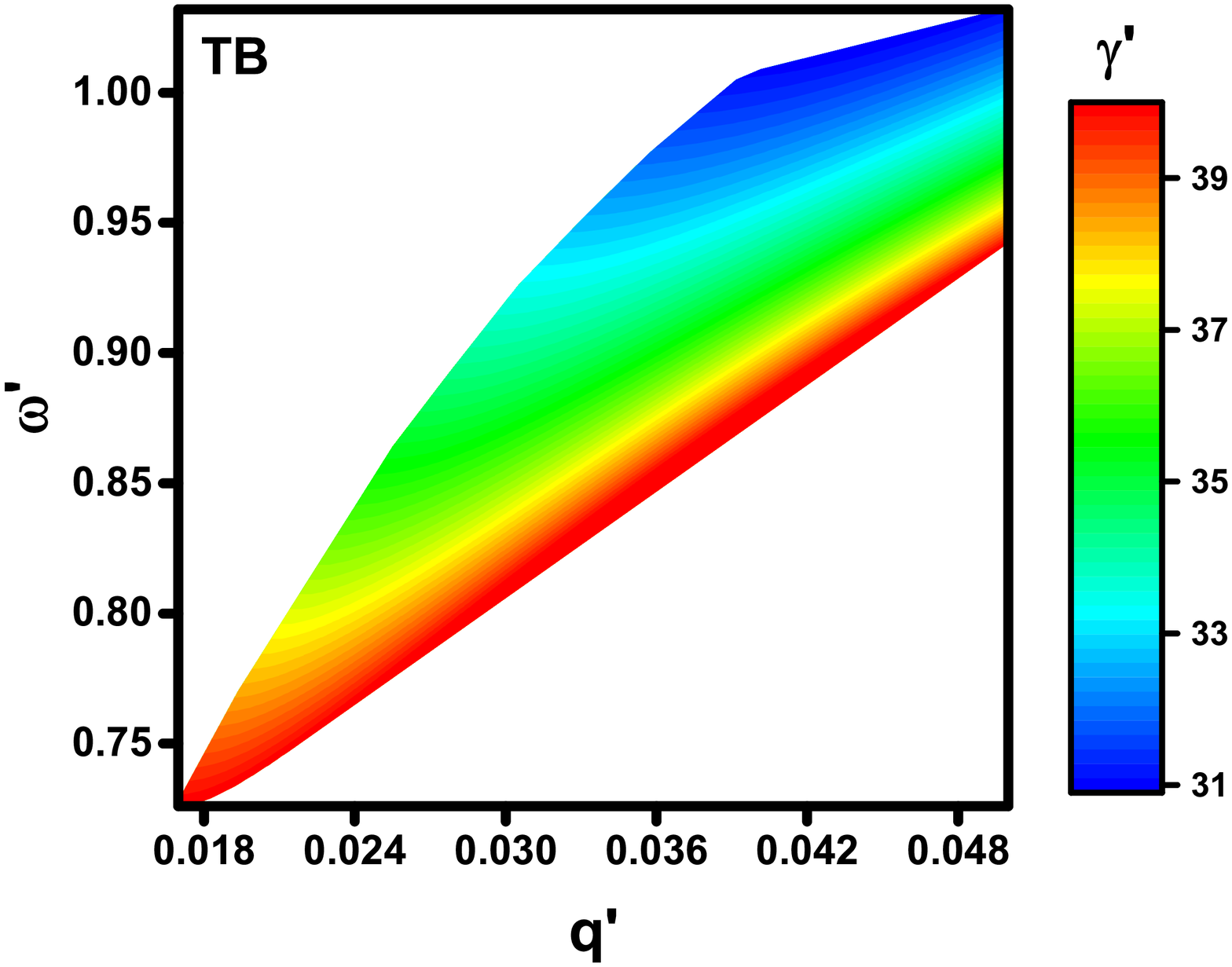}	
	\caption{\small{ SB and TB plasmon spectra for $\gamma'$ continuous. }}
	\label{fig:8}
\end{figure}

The plasmon spectrum in Figs. \ref{fig:5} and \ref{fig:7} involves only a few discrete values of $\gamma'$.  For a continuous $\gamma'$ we show it in Fig. \ref{fig:8} as a contour plot. Plasmons are not allowed outside the coloured regions.
For the same plasmon frequency and momentum, we  easily see  that the corresponding $\gamma'$  differ drastically. 

 \section{ Plasmons in a 2DEG}
For a 2DEG  the  single-particle wave function is given by 
\begin{equation}\label{e:26}
\psi_{\vec{k},s}(\vec{r})=\frac{1}{\sqrt{A}}e^{i\vec{k}.\vec{r}}X_s.
\end{equation}
Employing Eqs. \eqref{e:26} and \eqref{e:7} we obtain  the long wave-length limit $\chi_{non}^0$ in the form \cite{R-6}

\begin{equation}\label{e:27}
\chi_{non}^0(q',\omega')=\frac{n\hbar^2k_F^2}{mE_F^2}\frac{q'^2}{\omega'^2},
\end{equation}
where $n$ and $m$ are the charge density and electron mass, respectively. As for $\chi_{im}^0$,  
with Eqs \eqref{e:26} and \eqref{e:13}  and the  assumption that the  scattering is elastic and $\tau$  independent of the wave vector, we obtain 
\begin{equation}\label{e:28}
\chi_{im}^0(q',\omega')=\frac{k_F^2}{2\pi E_F}\,C(\omega',\gamma').
\end{equation}

\begin{figure}[t]
	\begin{center}	
		\includegraphics[height=4cm, width=4cm]{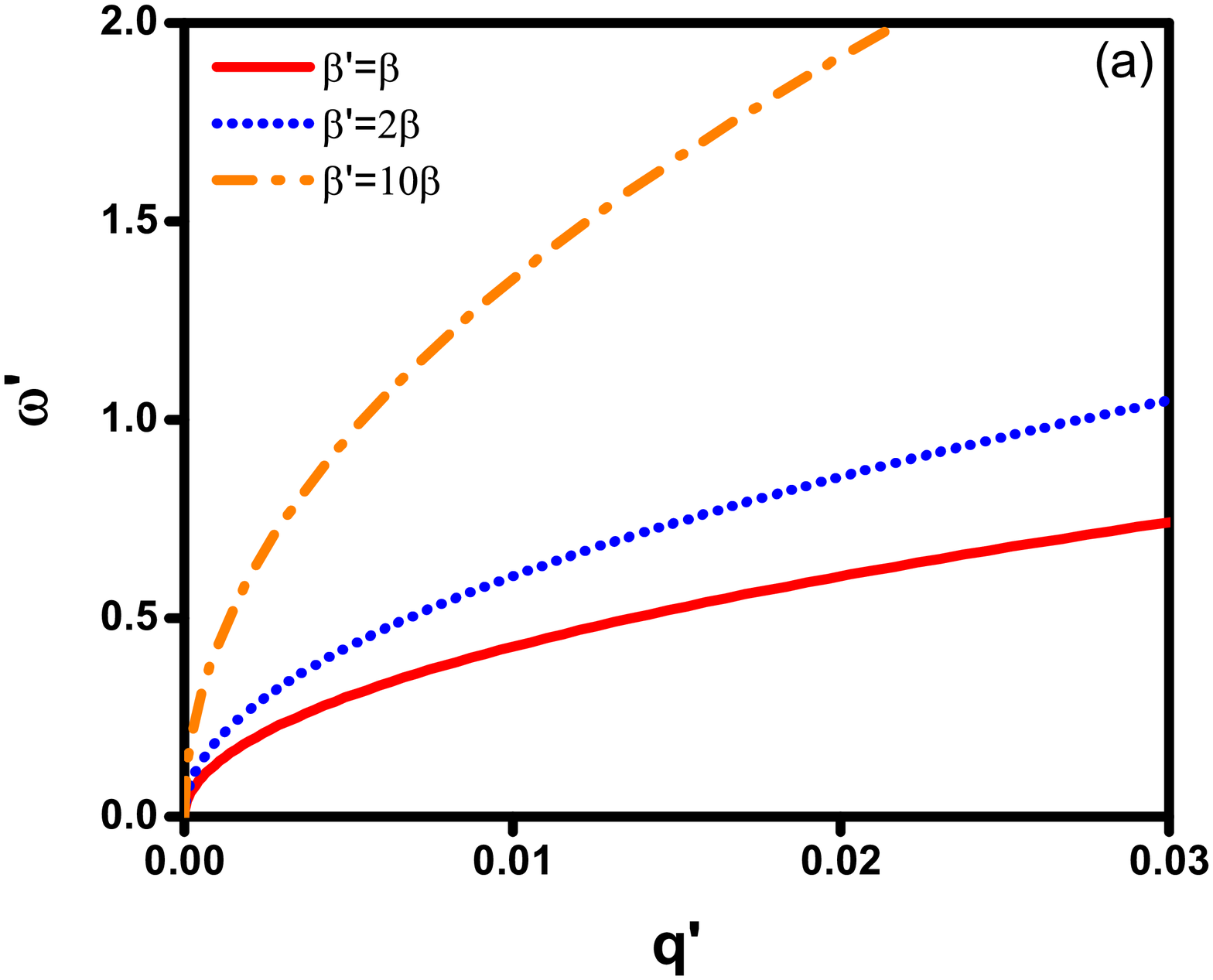}\quad
		\includegraphics[height=3.9cm, width=4cm]{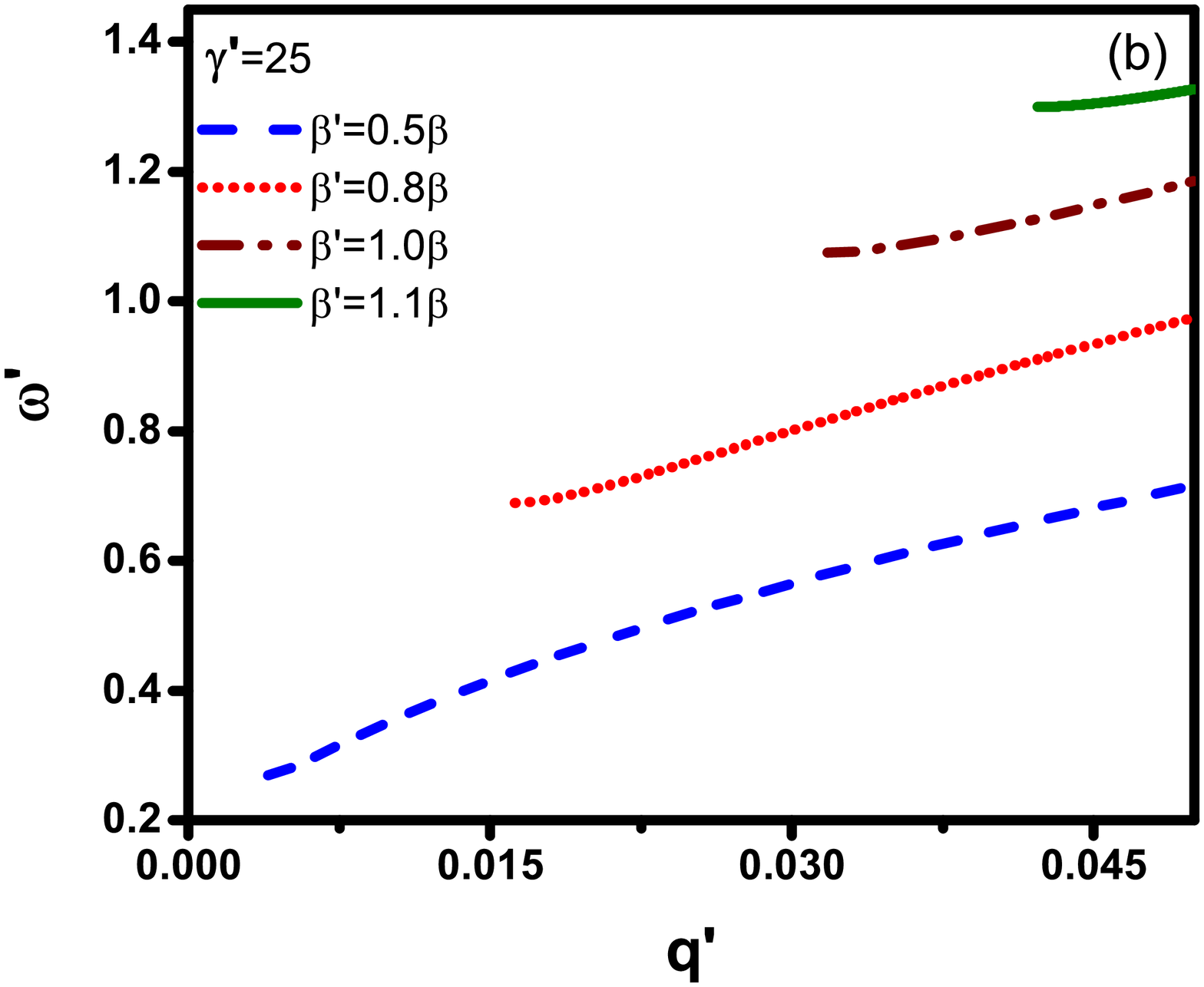}	
		\caption{\small{ 2DEG plasmon dispersion (a) in the absence and (b) presence of impurity scattering. In (b) $\gamma'=25$ is used. }}
	\label{fig:9}
	\end{center}
\end{figure}

In the long-wavelength limit the plasmon spectrum 
can be evaluated by utilizing  Eqs. \eqref{e:21}, \eqref{e:27}, and \eqref{e:28}. The result is  similar to graphene's plasmon spectrum, namely,

\begin{equation}\label{e:29}
q'=\frac{\omega'^2}{2\beta'}\left(1+\left[1-\frac{2\beta'^2\left(1-\delta_{\gamma',0}\right)}{\omega'\left(\omega'^2+\gamma'^2\right)}\right]^{1/2}\right)
\end{equation}
with $\beta'=4/k_F a_B$ and $a_B$  the Bohr radius. In  Fig. \ref{fig:9} we show this 2DEG  plasmon spectrum in the absence and presence of impurities
for various values of $\beta'$.  To contrast it  with that of graphene we give $\beta'$ in "units" of $\beta$. In  Fig. \ref{fig:9} (a) we can see  that for a fixed $q'$ the plasmon energy increases; this can explained as  follows. If the number of plasmon dipole-momenta increases we expect the energy of plasmon EM field to  increase as well. Note that in  Eq. \eqref{e:29} the plasmon momentum is inversely proportional to $\beta'$. In addition, $k_F$ in a 2DEG is proportional to the square root of the electron density, $k_F=\sqrt{2\pi n}$. Then one finds that  the plasmon momentum  $q'$ is likewise proportional to  $\sqrt{n}$.  In contrast, in graphene  the dimensionless plasmon momentum  is independent of the electron density $n$.
 In  Fig. \ref{fig:9} (b) we show the plasmon dispersion in a 2DEG in (a) the absence  and (b) the presence of impurity scattering; in (b) we took 
 $\gamma'=25$. It can be seen that  for  fixed plasmon momentum and decreasing $\beta'$ the plasmon energy decreases due to the reduction of the plasmon dipole momenta. Furthermore, compared to the graphene case $\beta'=\beta$, we can see that for the lower value of $\beta'$ the critical plasmon momentum becomes smaller due to the fact that scattering by impurities weakens with decreasing  electron density.  

\begin{figure}[t]
	\begin{center}		
		\includegraphics[height=4.0cm, width=6cm]{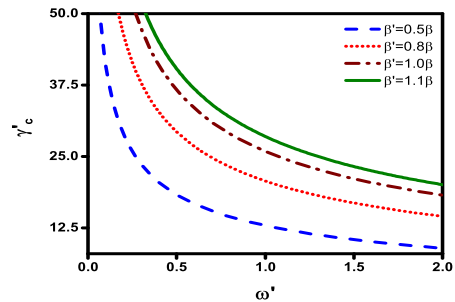} 	
		\caption{\small{ $\gamma^{'2DEG}_c$ versus $\omega'$ for several values of $\beta'$.}}
		\label{fig:10}
	\end{center}
\end{figure}

Further,  as in the case of  graphene, in a 2DEG the  critical $\gamma'$, below  which no plasmons are allowed, is obtained in the same way. It is given by 
\begin{equation}\label{e:30}
\gamma^{'2DEG}_c=\left[2\beta'^2/\omega'-\omega'^2\right]^{1/2}.
\end{equation}
We show it in  Fig. \ref{fig:10} $\gamma^{'2DEG}_c$ for several values of $\beta'$. It can be seen that for  fixed $\omega'$ and increasing 
$\beta'$ the value of $\gamma_c'$ increases as well. We further remark that, similar to graphene for SB transitions, with $\gamma_c'\neq 0$ fixed Eq. (33) allows an analytic evaluation of the allowed $\omega'$ which in turn determines the lower value of $q'$ below which no plasmons are allowed. 
One simply has to replace $\beta$ with $\beta'$ in Eq. (30) to obtain the corresponding $\omega_c$ and $q_c$.

\section{Summary}
We evaluated the linear-response function to an exter-\\nal stimulus, obtained an expression that is valid for elastiic scattering, and applied it to 
plasmons   in graphene and  the 2DEG in the random-phase approximation. This was achieved  by applying the van Hove limit to all operators and by utilizing appropriate super-operators  of the literature. The resulting  linear-response function, or DDRF, has two terms, $\chi_{non}$ which is independent of the scattering,   and  $\chi_{im}$ who does depend on it and produces results that are qualitatively and quantitatively  different from those of $\chi_{non}$. In graphene the term $\chi_{non}$ dominates the response in the long wavelength limit, i.e., for very low frequencies, while the term $\chi_{im}$ dominates  for all  other frequencies.

The main result of the term  $\chi_{im}$ is that introduces scattering-dependent wavevector limits below which no plasmons are allowed. It is also valid for all values of the wave vector, that is, it is not limited to the long-wavelength limit as $\chi_{non}$ is. Another nice feature  is that it simply explains and retrieves the Drude model results in the long-wavelength limit. We reported new plasmon results for i) graphene and ii) the 2DEG. In i) we distinguished between  intraband (SB) and  interband (TB) transitions. In both i) and ii) we obtained the scattering-induced  limits referred to above, analytical dispersion relations,   and their  well-known long-wavelength limit in the absence of scattering. An important difference between i) and ii) is that in the dimensionless units used the plasmon wave vector for graphene is independent of the electron density whereas in a 2DEG  it is proportional to its square root, $q'\propto \sqrt{n}$.  As discussed,  depending on the scattering strength $\gamma'$ the single-particle excitations due to scattering drastically modify  the frequency and wave vector domains ($\omega, q)$ of the collective  excitations.  The latter are suppressed below  a critical $\gamma'$.  

Finally, it's worth emphasizing that our results for both $\chi_{non}$ and the scattering-dependent $\chi_{im}$ are both fully {\it quantum mechanical.} This is different from the widespread practice of shifting the  frequency $\omega$ in the {\it classical} Drude result to $\omega+i\tau$ and relate $\tau$ to scattering. This is why our results  may appear strikingly different than the usual one. Similar results were obtained in \cite{R-7} for homogeneous systems, see Eq. (2.35) in there, but, as stated in Sec. I, the corresponding $\chi_{im}$ term was not explicitly evaluated.

\section*{Acknowledgments}
This work was supported by the  Canadian NSERC Grant No. OGP0121756

\noindent Electronic addresses: $^{\dag}$m\_bahra@live.concordia.com, \\$^{\ddag}$p.vasilopoulos@concordia.ca
\appendix{}

\appendix{}

\section{Scattering-dependent part of the DDRF
}
In order to simplify derivation we rewrite Eq. (\ref{e:9}) as 
\begin{equation}\label{A:1}
c_l(t)=c_l^{im}(t)+c_l^{non}(t),
\end{equation}
with
\begin{equation}\label{A:2}
c_l^{im}(t)=e^{-\Lambda_l t}c_l^d  
\end{equation}
and
\begin{equation}
c_l^{non}(t)=e^{iH_0t/\hbar}c_le^{-iH_0t/\hbar}. 
\end{equation}
Notice that $c_l^{non}(t)$ is the time evolution of the operator in the absence of impurity. There is  an advantage  in splitting time evolution of operator 
as in Eq. (35): applying it to the density operator we have 
\begin{equation}\label{A:3}
\rho(\vec{r},t)=\rho_{im}(\vec{r},t)+\rho_{non}(\vec{r},t). 
\end{equation}
Substituting (\ref{A:3} ) into expression  (\ref{e:4}) and   exploiting the  algebra \cite{R-1}
\begin{equation}
\Big\langle \left[\rho_{im}(\vec{r},t),\rho_{non}(\vec{r^{'}},t')\right] \Big\rangle_0=0 
\end{equation}
Eq. (\ref{e:4}) gives 
\begin{equation}
\chi^0(\vec{r},\vec{r^{'}},t)=\chi_{im}^0(\vec{r},\vec{r^{'}},t)+\chi^0_{non}(\vec{r},\vec{r^{'}},t),
\end{equation}
where  $\chi^0_{non}$ is  represented in the frequency domain by Eq. (\ref{e:7}). The result for $\chi_{im}^0$ is 
\begin{equation}\label{A:7}
\chi_{im}^0(\vec{r},\vec{r^{'}},t)=	-\frac{i}{\hbar}\Theta(t)\Big\langle \left[\rho_{im}(\vec{r},t),\rho_{im}(\vec{r^{'}})\right]\Big\rangle_0 .	   
\end{equation}
Using Eqs. \eqref{A:2} and \eqref{A:3}, Eq. \eqref{A:7},  and integrating over time we  obtain Eq. (\ref{A:7}) in the frequency domain  as 
\begin{eqnarray}\label{A:8}
&&\chi_{im}^0(\vec{r},\vec{r^{'}},\omega)=	-\frac{i}{\hbar}\int_{-\infty}^{\infty}dt\,\Theta(t)e^{i\omega t}\big\langle e^{-\Lambda_{ij} t}\big\rangle_b 
\nonumber\\*
&&\hspace*{-0.7cm}\times\sum_{i,j}\sum_{m,n}\phi_i^{*}(\vec{r})\phi_j(\vec{r}) \phi_m^{*}(\vec{r^{'}})\phi_n(\vec{r^{'}}) (f_i-f_j) \delta_{i,n}\delta_{j,m},
\end{eqnarray}
where we defined  $\Lambda_{ij}\equiv \Lambda_i+\Lambda_j.$
Evaluating the integral in Eq. \eqref{A:8}we obtain Eq. \eqref{e:14-1}.


\end{document}